\documentclass{article}

\usepackage{arxiv,times}

\usepackage[utf8]{inputenc} %
\usepackage[T1]{fontenc}    %
\usepackage{url}            %
\usepackage{booktabs}       %
\usepackage{amsfonts}       %
\usepackage{nicefrac}       %
\usepackage{microtype}      %
\usepackage{url}
\usepackage{mathtools}
\usepackage{graphicx}
\usepackage{dcolumn}
\newcolumntype{d}[1]{D{.}{.}{#1}}
\usepackage{float}
\usepackage{subfig}
\usepackage{xcolor}

\usepackage{quantikz}
\usepackage{physics}
\usepackage{amsmath}
\usepackage{amssymb}
\usepackage{textcomp}
\usepackage{amsthm}
\numberwithin{equation}{section}
\usepackage{slashed}
\usepackage{braket}
\usepackage{enumitem}
\usepackage{dsfont}
\usepackage{bbm}
\usepackage{bm}
\usepackage{hyperref}
\hypersetup{colorlinks=true, citecolor=mypurple,linkcolor=firebrick,urlcolor=firebrick}
\usepackage[nameinlink,capitalize,noabbrev]{cleveref}

\usepackage[bottom]{footmisc}
\usepackage{dashbox}
\usepackage{colortbl}
\usepackage{booktabs}
\usepackage{multirow}

\usepackage{mdframed}
\usepackage[many]{tcolorbox}
\tcbuselibrary{breakable}

\usepackage{tikz}
\usetikzlibrary{decorations.pathreplacing, fit}
\usepackage{titletoc}
\usepackage{titlesec}
\titlespacing\section{0pt}{4pt plus 4pt minus 2pt}{-2pt plus 2pt minus 2pt}
\titlespacing\subsection{0pt}{2pt plus 4pt minus 2pt}{-2pt plus 2pt minus 2pt}
\titlespacing\subsubsection{0pt}{2pt plus 4pt minus 2pt}{-2pt plus 2pt minus 2pt}

\def\ceil#1{\lceil #1 \rceil}

\def\1{\bm{1}}

\def\eps{{\epsilon}}

\DeclareMathAlphabet{\mathsfit}{\encodingdefault}{\sfdefault}{m}{sl}
\SetMathAlphabet{\mathsfit}{bold}{\encodingdefault}{\sfdefault}{bx}{n}

\let\hat\widehat

\newcommand{\calL}{\mathcal{L}}

\newcommand{\calN}{\mathcal{N}}

\newcommand{\R}{\mathbb{R}}
\newcommand{\C}{\mathbb{C}}

\DeclareMathOperator*{\argmax}{arg\,max}
\DeclareMathOperator*{\argmin}{arg\,min}

\newcommand{\mainloss}{\ell_{\text{MT}}}
\newcommand{\ssloss}{\ell_{\text{AE}}}
\newcommand{\layerE}{L_{\text{E}}}
\newcommand{\layerD}{L_{\text{D}}}
\newcommand{\layerM}{L_{\text{M}}}
\newcommand{\Ntest}{N_{\text{Testing}}}
\newcommand{\Ntrain}{N_{\text{Training}}}

\makeatletter
\def\th@remark{%
  \thm@headfont{\bfseries}%
  \normalfont %
  \thm@preskip\topsep \divide\thm@preskip\tw@
  \thm@postskip\thm@preskip
}
\makeatother

\theoremstyle{definition}
\newtheorem{theorem}{Theorem}[section]
\tcolorboxenvironment{theorem}{
  breakable,
  colback=fundamental!10,
  colframe=white,%
  width=\dimexpr\linewidth+10pt\relax,%
  enlarge left by=-5pt,%
  enlarge right by=-5pt,%
  boxsep=5pt,%
  boxrule=0pt,
  left=0pt,right=0pt,top=0pt,bottom=0pt,
  sharp corners,
  before skip=\topsep,
  after skip=\topsep
}

\tcolorboxenvironment{lemma}{
  breakable,
  colback=fundamental!10,
  colframe=white,%
  width=\dimexpr\linewidth+10pt\relax,%
  enlarge left by=-5pt,%
  enlarge right by=-5pt,%
  boxsep=5pt,%
  boxrule=0pt,
  left=0pt,right=0pt,top=0pt,bottom=0pt,
  sharp corners,
  before skip=\topsep,
  after skip=\topsep
}

\tcolorboxenvironment{corollary}{
  breakable,
  colback=fundamental!10,
  colframe=white,%
  width=\dimexpr\linewidth+10pt\relax,%
  enlarge left by=-5pt,%
  enlarge right by=-5pt,%
  boxsep=5pt,%
  boxrule=0pt,
  left=0pt,right=0pt,top=0pt,bottom=0pt,
  sharp corners,
  before skip=\topsep,
  after skip=\topsep
}

\tcolorboxenvironment{proposition}{
  breakable,
  colback=fundamental!10,
  colframe=white,%
  width=\dimexpr\linewidth+10pt\relax,%
  enlarge left by=-5pt,%
  enlarge right by=-5pt,%
  boxsep=5pt,%
  boxrule=0pt,
  left=0pt,right=0pt,top=0pt,bottom=0pt,
  sharp corners,
  before skip=\topsep,
  after skip=\topsep
}
\theoremstyle{definition}

\tcolorboxenvironment{definition}{
  breakable,
  colback=fundamental!10,
  colframe=white,%
  width=\dimexpr\linewidth+10pt\relax,%
  enlarge left by=-5pt,%
  enlarge right by=-5pt,%
  boxsep=5pt,%
  boxrule=0pt,
  left=0pt,right=0pt,top=0pt,bottom=0pt,
  sharp corners,
  before skip=\topsep,
  after skip=\topsep
}
\newtheorem{remark}{Remark}[section]

\tcolorboxenvironment{assumption}{
  breakable,
  colback=fundamental!10,
  colframe=white,%
  width=\dimexpr\linewidth+10pt\relax,%
  enlarge left by=-5pt,%
  enlarge right by=-5pt,%
  boxsep=5pt,%
  boxrule=0pt,
  left=0pt,right=0pt,top=0pt,bottom=0pt,
  sharp corners,
  before skip=\topsep,
  after skip=\topsep
}

\tcolorboxenvironment{claim}{
  breakable,
  colback=fundamental!10,
  colframe=white,%
  width=\dimexpr\linewidth+10pt\relax,%
  enlarge left by=-5pt,%
  enlarge right by=-5pt,%
  boxsep=5pt,%
  boxrule=0pt,
  left=0pt,right=0pt,top=0pt,bottom=0pt,
  sharp corners,
  before skip=\topsep,
  after skip=\topsep
}

\tcolorboxenvironment{problem}{
  breakable,
  colback=fundamental!10,
  colframe=white,%
  width=\dimexpr\linewidth+10pt\relax,%
  enlarge left by=-5pt,%
  enlarge right by=-5pt,%
  boxsep=5pt,%
  boxrule=0pt,
  left=0pt,right=0pt,top=0pt,bottom=0pt,
  sharp corners,
  before skip=\topsep,
  after skip=\topsep
}

\tcolorboxenvironment{question}{
  breakable,
  colback=fundamental!10,
  colframe=white,%
  width=\dimexpr\linewidth+10pt\relax,%
  enlarge left by=0pt,%
  enlarge right by=0pt,%
  boxsep=5pt,%
  boxrule=0pt,
  left=0pt,right=0pt,top=0pt,bottom=0pt,
  sharp corners,
  before skip=\topsep,
  after skip=\topsep
}
\crefname{question}{Question}{Questions}

\crefname{theorem}{Theorem}{Theorems}
\crefname{proposition}{Proposition}{Propositions}
\crefname{lemma}{Lemma}{Lemmas}
\crefname{corollary}{Corollary}{Corollaries}
\crefname{definition}{Definition}{Definitions}
\crefname{assumption}{Assumption}{Assumptions}
\crefname{remark}{Remark}{Remarks}
\crefname{problem}{Problem}{Problems}
\crefname{property}{Property}{property}

\numberwithin{equation}{section}
\numberwithin{theorem}{section}
\numberwithin{proposition}{section}
\numberwithin{definition}{section}
\numberwithin{lemma}{section}
\numberwithin{assumption}{section}
\numberwithin{remark}{section}

\definecolor{lightyellow}{rgb}{1.0, 0.95, 0.7}
\definecolor{Blue}{rgb}{0, 0, 0.8}
\definecolor{blue}{rgb}{0,0,1}
\definecolor{mydarkblue}{rgb}{0,0.08,0.45}
\definecolor{mydarkblue2}{rgb}{0.133, 0.133, 0.698}
\definecolor{echodrk}{HTML}{0099cc}
\definecolor{mymauve}{rgb}{0.58,0,0.82}
\definecolor{darkgreen}{rgb}{0,0.40,0}
\definecolor{firebrick}{rgb}{0.698,0.133,0.133}
\definecolor{coolteal}{rgb}{0, 0.45, 0.45}
\definecolor{olive}{rgb}{0.1, 0.3, 0}
\definecolor{mypurple}{rgb}{0.5,0,0.5}
\definecolor{almond}{rgb}{0.94, 0.87, 0.8}

\definecolor{blue_ampEncoding}{HTML}{DAE8FC}
\definecolor{green_encoder}{HTML}{D5E8D4}
\definecolor{purple_decoder}{HTML}{E1D5E7}
\definecolor{yellow_measure}{HTML}{FFF2CC}
\definecolor{gray_block}{HTML}{F5F5F5}
\definecolor{pink_dru}{HTML}{FAD9D5}
\definecolor{orange_v}{HTML}{FAD7AC}

\definecolor{colorA}{rgb}{1,0,0}
\definecolor{colorB}{rgb}{0,0.3,1}
\definecolor{colorC}{rgb}{0.9,0.8,0.2}
\definecolor{colorD}{rgb}{0,0.65,0}
\definecolor{lesslightgray}{rgb}{0.5,0.5,0.5}
\definecolor{light-gray}{gray}{0.95}
\definecolor{lightblue}{rgb}{0.90, 0.95, 1.0}
\definecolor{fundamental}{RGB}{55, 110, 111}

\newcommand*{\annot}[1]{\tag*{\footnotesize{\textcolor{black!50}{\big(#1\big)}}}}

\makeatletter
\let\save@mathaccent\mathaccent
\newcommand*\if@single[3]{%
    \setbox0\hbox{${\mathaccent"0362{#1}}^H$}%
    \setbox2\hbox{${\mathaccent"0362{\kern0pt#1}}^H$}%
    \ifdim\ht0=\ht2 #3\else #2\fi
}
\newcommand*\rel@kern[1]{\kern#1\dimexpr\macc@kerna}
\newcommand*\widebar[1]{\@ifnextchar^{{\wide@bar{#1}{0}}}{\wide@bar{#1}{1}}}
\newcommand*\wide@bar[2]{\if@single{#1}{\wide@bar@{#1}{#2}{1}}{\wide@bar@{#1}{#2}{2}}}
\newcommand*\wide@bar@[3]{%
    \begingroup
    \def\mathaccent##1##2{%
        \let\mathaccent\save@mathaccent
        \if#32 \let\macc@nucleus\first@char \fi
        \setbox\z@\hbox{$\macc@style{\macc@nucleus}_{}$}%
        \setbox\tw@\hbox{$\macc@style{\macc@nucleus}{}_{}$}%
        \dimen@\wd\tw@
        \advance\dimen@-\wd\z@
        \divide\dimen@ 3
        \@tempdima\wd\tw@
        \advance\@tempdima-\scriptspace
        \divide\@tempdima 10
        \advance\dimen@-\@tempdima
        \ifdim\dimen@>\z@ \dimen@0pt\fi
        \rel@kern{0.6}\kern-\dimen@
        \if#31
        \overline{\rel@kern{-0.6}\kern\dimen@\macc@nucleus\rel@kern{0.4}\kern\dimen@}%
        \advance\dimen@0.4\dimexpr\macc@kerna
        \let\final@kern#2%
        \ifdim\dimen@<\z@ \let\final@kern1\fi
        \if\final@kern1 \kern-\dimen@\fi
        \else
        \overline{\rel@kern{-0.6}\kern\dimen@#1}%
        \fi
    }%
    \macc@depth\@ne
    \let\math@bgroup\@empty \let\math@egroup\macc@set@skewchar
    \mathsurround\z@ \frozen@everymath{\mathgroup\macc@group\relax}%
    \macc@set@skewchar\relax
    \let\mathaccentV\macc@nested@a
    \if#31
    \macc@nested@a\relax111{#1}%
    \else
    \def\gobble@till@marker##1\endmarker{}%
    \futurelet\first@char\gobble@till@marker#1\endmarker
    \ifcat\noexpand\first@char A\else
    \def\first@char{}%
    \fi
    \macc@nested@a\relax111{\first@char}%
    \fi
    \endgroup
    }
\makeatother

\title{Test-Time Training with Quantum Auto-Encoder: From Distribution Shift to Noisy Quantum Circuits}

\author{%
Damien Jian$^{1,}$\footnotemark[1]
\quad
Yu-Chao Huang$^{2,4,}$\footnotemark[1] 
\quad 
Hsi-Sheng Goan$^{2,3,4}$ \\
$^1$ Graduate Institute of Applied Physics, \\
National Taiwan University, Taipei 106319, Taiwan\\
$^2$ Department of Physics and Center for Theoretical Physics, \\
National Taiwan University, Taipei 106319, Taiwan\\ 
$^3$ Center for Quantum Science and Engineering, \\National Taiwan University, Taipei 106319, Taiwan\\ 
$^4$ Physics Division, National Center for Theoretical Sciences, Taipei 106319, Taiwan
}

\iclrfinalcopy %
\begin{document}

\renewcommand{\thefootnote}{\fnsymbol{footnote}}
\footnotetext[1]{Equal contribution. Authors are listed in reverse alphabetical order.}
\footnotetext[2]{Correspondence to: Yu-Chao Huang <yuchaohuang@g.ntu.edu.tw>.}
\footnotetext[3]{Project Website: \url{https://physics-morris.github.io/QTTT.html}.}
\maketitle

\begin{abstract}
In this paper, we propose test-time training with the quantum auto-encoder (QTTT).
QTTT adapts to (1) data distribution shifts between training and testing data and (2) quantum circuit error by minimizing the self-supervised loss of the quantum auto-encoder.
Empirically, we show that QTTT is robust against data distribution shifts and effective in mitigating random unitary noise in the quantum circuits during the inference.
Additionally, we establish the theoretical performance guarantee of the QTTT architecture.
Our novel framework presents a significant advancement in developing quantum neural networks for future real-world applications and functions as a plug-and-play extension for quantum machine learning models.

\begin{figure}[htp!]
    \centering
    \includegraphics[width=0.7\linewidth]{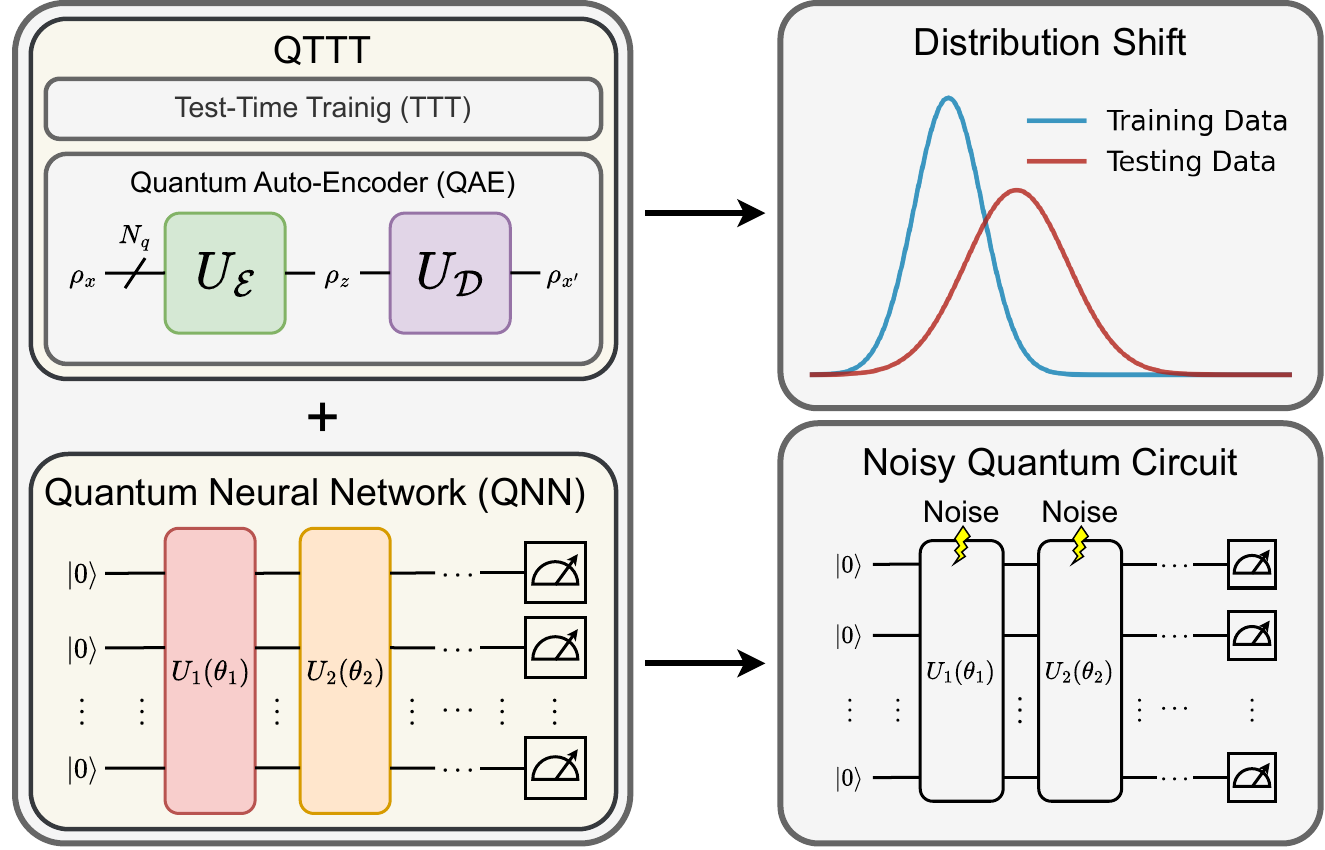}
    \vspace{0.0em}
    \caption{
    \textbf{Test-Time Training with Quantum Auto-Encoder (QTTT) as a Flexible Plug-and-Play Extension for Quantum Neural Networks (QNNs).}
    QTTT adapts to distribution shift and mitigates random unitary noise present during inference by minimizing the self-supervised loss from the quantum auto-encoder.
    QTTT integrates easily with existing QNNs, with a mild computational overhead ratio during training and test-time training that is inversely proportional to the main QNN depth. 
    QTTT serves as a bridge toward the practical utility of QNNs in the future.
    }
    \label{fig:cover}
\end{figure}

\end{abstract}

\clearpage

\section{Introduction}\label{sec:intro}

Quantum computing offers an alternative approach to computation from classical computers, grounded in quantum mechanics principles such as superposition, entanglement, and interference, with the potential for reshaping the future of computing~\citep{herrmann2022realizing,Shor_1997}.
In the noisy intermediate-scale quantum (NISQ) era, quantum machine learning (QML), as an important component of variational quantum algorithms (VQAs), offers a fruitful playground for exploring theoretical aspects such as expressive power~\citep{wen2024enhancing}, generalization~\citep{abbas2021power,huang2021power}, and overparameterization~\citep{larocca2023theory}.
Additionally, it drives methodological innovation by leveraging quantum mechanical principles for learning quantum data and error correction scheme~\citep{herrmann2022realizing,cong2019quantum}.

Supervised learning makes predictions based on labeled training datasets, and generalizes to make predictions on the testing, unseen data.
However, the trained model is fixed during deployment.
Even minor variations in the testing data distribution compared to the training data, or unexpected sources of noise in the quantum circuit during inference, such as decoherence, gate errors, or readout inaccuracies, can jeopardize the model’s generalization abilities.

Classical machine learning community proposes test-time training (TTT), an on-the-fly adaption to distribution shift in the testing data~\citep{sun2024learning,gandelsman2022test,sun2020test}.
TTT updates the model parameters by minimizing self-supervised loss with gradient descent during test time.
The ``overfitting'' or in-context learning approach on testing data has demonstrated its effectiveness in recent innovations with large language models (LLMs)~\citep{snell2024scaling}, where a larger portion of computational resources is allocated during inference.

We propose test-time training with a quantum auto-encoder (QTTT) to tackle two key challenges: (1) enhancing model generalization when the distribution of training and testing data differ, and (2) enabling noise-awareness for deployment scenarios where the QML model faces noise not presenting during the training stage.
QTTT deploys a Y-shaped architecture following \citet{gandelsman2022test, sun2020test}.
The architecture comprises a parameter-shared quantum encoder, a quantum decoder, and the main QML model for downstream tasks. 
During training, the multi-task objective is minimized on both the quantum auto-encoder (QAE) loss function and the classification loss function from the main QML model.
At test time, the shared quantum encoder is trained on a self-supervised task by minimizing the QAE loss function for quantum state recovery. 
This approach leverages the idea of self-supervision to capture the underlying distribution of testing data. 
Moreover, our experimental results show that minimizing the self-supervised loss during testing effectively mitigates noise in quantum circuits that may not be present during training.

The QTTT framework (see \cref{fig:cover}) provides a plug-and-play extension for existing QML models, addressing the challenges related to the distribution shift from training and testing data and quantum circuit error during the deployment stage. 
This innovative framework brings QML closer to real-world applications, accommodating deployment scenarios where quantum computers may exhibit different noise characteristics than those encountered during training.

\paragraph{Contributions.} We summarize our main contribution as follows:
\begin{itemize}
    \item We propose QTTT, a novel test-time training framework for QNNs, offering a plug-and-play architecture extension to address distribution shifts between training and testing data and random unitary noise during inference.
    \item We theoretically analyze the performance guarantee of QTTT by examining the main task loss under gradient descent during test time while optimizing the shared parameters. 
    Additionally, we present the computation overhead ratio of QTTT relative to training, which reduces when the depth of the main task model increases.
    \item We conduct comprehensive experiments and show that (1) QTTT effectively improves the performance of QNNs on corrupted testing data and (2) QTTT robustly mitigates random unitary noises in quantum circuits.
\end{itemize}

\paragraph{Organizations.}
We organize the paper as follows.
In \cref{sec:related}, we organize related works of test-time training, QNNs, and noise-aware variational quantum algorithms.
We introduce the QTTT method in \cref{sec:method} and provide theoretical analysis on its performance guarantee in \cref{sec:theory}.
In \cref{sec:exp}, we demonstrate the effectiveness of QTTT on addressing distribution shifts between training and testing data and mitigating quantum circuit error.
Finally, we discuss the future direction and limitation of QTTT in \cref{sec:lim} and conclude this paper in \cref{sec:conclusion}.

\paragraph{Notations.}
We denote scalars and vectors in lowercase letters and matrices in uppercase letters,
$\|{x}\|$ the \(\ell_2\)-norm of vector $x$.
We use $[\cdot, \ldots, \cdot]$ to denote the concatenation operation.
The inner product of vectors $x$ and $y$ of the same dimension is denoted by $\langle x, y \rangle$.

\section{Related Works}\label{sec:related}

\paragraph{Quantum Machine Learning.} 
QML is often categorized into three main types~\citep{bowles2024better}: quantum neural networks, convolutional quantum neural networks, and quantum kernel methods. 
QNN utilizes parameterized quantum circuits (PQCs) to encode data and evolve the quantum state, with parameter training carried out through classical optimization. 
Following \citet{bowles2024better}, popular QNN models include the data reuploading classifier~\citep{perez2020data}, dressed quantum circuits~\citep{mari2020transfer}, and the quantum Boltzmann machine~\citep{amin2018quantum}.
On the other hand, quantum kernel methods leverage quantum computers to evaluate kernel functions, embedding classical data into quantum states to measure similarity between data points through inner products~\citep{bowles2024better}. 
Following \citet{bowles2024better}, popular quantum kernel methods include the IQP kernel classifier~\citep{havlivcek2019supervised} and the projected quantum kernel~\citep{huang2021power}.
Finally, quantum convolutional neural networks (QCNNs) is the quantum analog of classical convolutional neural networks, incorporating inductive biases of translation invariance and locality. 
Examples of such models include Quantum Convolutional Neural Networks~\citep{cong2019quantum} and WeiNet~\citep{wei2022quantum}.

\paragraph{Test-Time Training.}
The idea of TTT, proposed by \citet{sun2020test}, minimizes the self-supervised loss of a four-way rotation prediction task during test time to address distribution shift while preserving the model's original performance.
Later, they enhanced performance by employing a masked auto-encoder as the self-supervised task for object recognition~\citep{gandelsman2022test}. 
Both \citet{gandelsman2022test,sun2020test} consider testing data arriving sequentially, framing TTT as a one-sample learning problem. 
Conversely, other studies have explored TTT in various applications using either the entire dataset or batches from the testing distribution~\citep{liu2021ttt,wang2020tent}.
Additionally, the idea of test-time training is also employed in LLMs~\citep{snell2024scaling} and video stream~\citep{wang2023test}.

\section{Methodology}\label{sec:method}

In this section, we first formulate the problem of (1) distribution shift between training and testing data and (2) quantum circuit errors during inference that are absent during training in \cref{sec:problem}.
In \cref{sec:arch}, we present the architecture of QTTT and various components.

\subsection{Problem Formulation}\label{sec:problem}

\paragraph{Classification Problem.}
In an classification problem with $C$ class, given a feature vector $x\in\R^{ d_x }$ of dimension $d_x$ and its corresponding one-hot label vector $y\in\R^{C}$, a classifier $g_\theta:\R^{ d_x }\to\R^{C}$ with trainable parameters $\theta$ predicts the class of the input, i.e., $\Tilde{y} = g_{\theta}(x)$. We use $\ell(x, y; \theta)$ to denote the loss of the model given data pair $(x, y)$.
Given training dataset $D_\text{train}\coloneqq\{(x, y)_i\}$ with number of data point $N_\text{train}$, drawn from data distribution $P$, the model is obtained by minimizing the empirical loss,
\begin{align*}
    \calL_\text{train}(\theta) &= \frac{1}{N_\text{train}}\sum_{(x, y) \in D_\text{train}} \ell(x, y; \theta), 
    &
    \theta^\star = \argmin_{\theta}\calL_\text{train}(\theta).
\end{align*}

\paragraph{Distribution Shift.}
Given testing data $D_\text{test}\coloneqq\{(x, y)_i\}$ with number of data point $N_\text{test}$, drawn from a different data distribution $P'$, the empirical loss is
\begin{align*}
    \calL_\text{test}(\theta) = \frac{1}{N_\text{test}}\sum_{(x, y) \in D_\text{test}} \ell(x, y; \theta). 
\end{align*}
We use QTTT to update the parameter set $\theta^{^\circledast}$ by performing gradient descent starting from $\theta^{\star}$ such that $ \calL_\text{test}(\theta^{^\circledast}) < \calL_\text{test}(\theta^\star)$.

\paragraph{Noisy Qunatum Cirucit.}
Given testing data $D_\text{test}\coloneqq\{(x, y)_j\}$ drawn from data distribution $P$, we consider a noisy classifier $g^{\prime}_\theta$.
Then use QTTT to update the parameter set $\theta^{^\circledast}$ from $\theta^{\star}$.

\subsection{QTTT Architecture}\label{sec:arch}
\begin{figure}[htp!]
    \centering
    \includegraphics[width=1.0\linewidth]{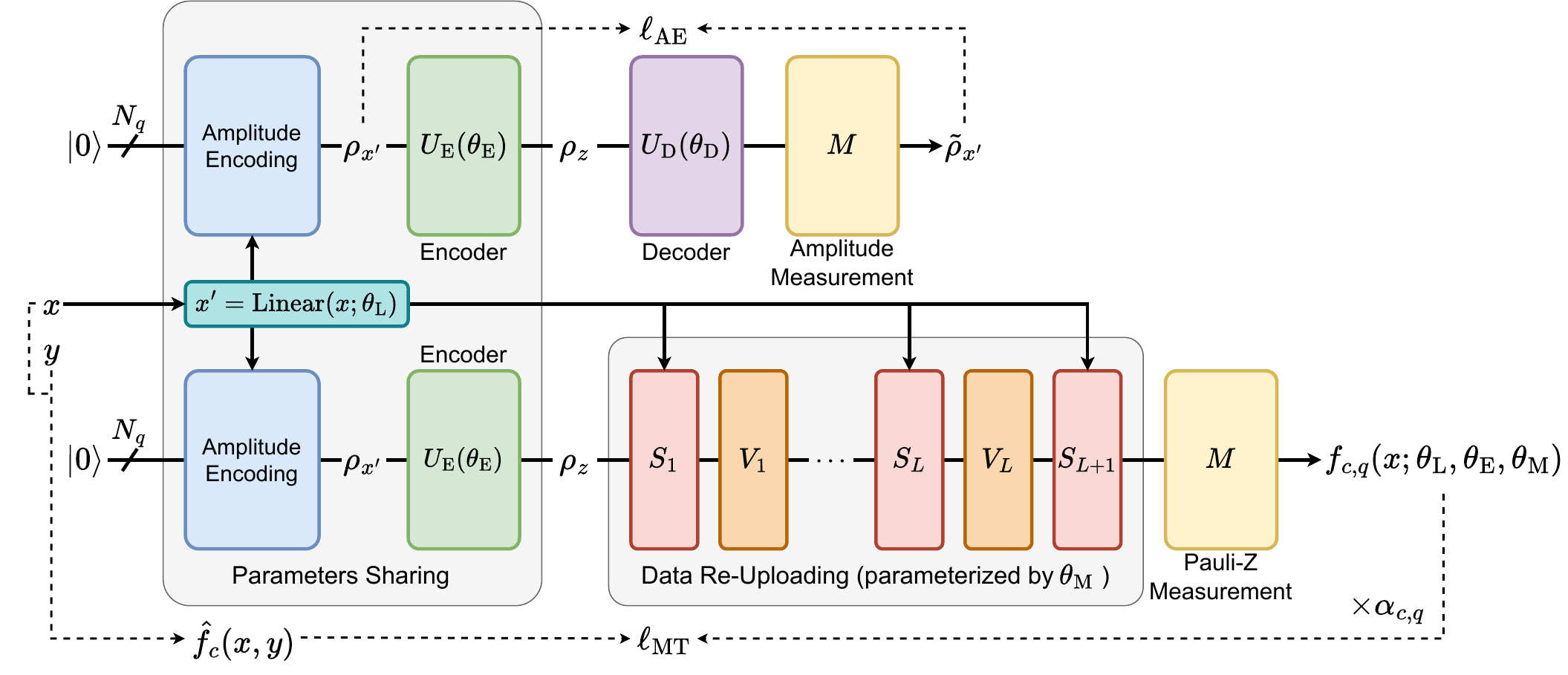}
    \vspace{-1.0em}
    \caption{
    \textbf{The Overall Architecture of Test-Time Training with Quantum Auto-Encoder (QTTT).}
    QTTT features a Y-shaped QNN architecture.
    The main branch is a data re-uploading classifier for classification tasks, with the corresponding loss function $\mainloss (\cdot; \theta_{\text{L}}, \theta_{\text{E}}, \theta_{\text{M}})$.
    The auxiliary branch is a QAE, with the corresponding loss function $\ssloss (\cdot; \theta_{\text{L}}, \theta_{\text{E}}, \theta_{\text{D}})$.
    We use a multi-task objective during training, with trainable parameters $(\theta_{\text{L}}, \theta_{\text{E}}, \theta_{\text{D}}, \theta_{\text{M}})$.
    In the test-time training stage, the shared parameters $(\theta_{\text{L}}, \theta_{\text{E}})$ in the linear layer and encoder enable QTTT to capture distribution shifts in test data or quantum circuit errors.
    }
    \label{fig:arch}
\end{figure}

The architecture is designed to have two branches after a shared encoding block.
The first branch is the decoding block, forming a quantum auto-encoder (QAE) with a shared quantum encoding block.
The second branch is the main classification task block. \cref{fig:arch} shows the architecture and the calculation flow of QTTT.

\paragraph{Pre-Processing of Data.} 
To enhance the QTTT effectiveness, we first feed data $x$ into a linear layer which outputs a feature vector $x'\in\R^{ d_x }$ which will also be used many times in data re-uploading part in the classification block later, the other branch.
\begin{align*}
x'=Mx+b = \text{Linear}(x; \theta_\text{L}),
\end{align*}
where $\theta_\text{L} \coloneqq [\text{Flatten}(M), b]\in\R^{ d_x ( d_x +1)}$, $M\in\R^{ d_x \times  d_x }$, and $b\in\R^{ d_x }$ are the parameters in the linear layer, where $\text{Flatten}(\cdot)$ means flattening the input to a row vector. 
Then we use amplitude encoding ~\citep{mottonen2004transformation} to encode the classical data $x'$ into quantum data $\rho_{x'}\in\C^{2^{N_q}\times2^{N_q}}$ into the quantum circuit where $N_q$ is the number of qubits of the quantum circuit,
\begin{align*}
    \rho_{x'} = \text{AmpEncoding}(x') = U_\text{A}(x')\rho_0 U_\text{A}^\dagger(x'),
\end{align*}
where $\rho_0$ is the initial state of the muti-qubit quantum circuit.
\paragraph{Auto-Encoder Branch.} The encoding and decoding block are denoted as parameterized unitary operators $U_\text{E}({\theta}_\text{E})$ and $U_\text{D}({\theta}_\text{D})$ with trainable parameters ${\theta}_\text{E}$ and ${\theta}_\text{D}$, respectively.
The encoding block transforms the quantum state $\rho_{x'}$ in data space to a latent variable $\rho_z$ in the latent space, and the decoding block sends the latent variable back to the data space $\Tilde{\rho}_{x'}$~\citep{romero2017quantum},
\begin{align*}
    {\rho_z}
    &= U_\text{E}({\theta}_\text{E}) {\rho_{x'}}U_\text{E}^\dagger({\theta}_\text{E}), &
    {\Tilde{\rho}_{x'}} &= U_\text{D}({\theta}_\text{D}) {\rho_z} U_\text{D}^\dagger({\theta}_\text{D}).
\end{align*}
The QAE loss $\ell_\text{AE}$ is defined as the relative infidelity between the original data state $\rho_{x'}$ and the recovered data state $\Tilde{\rho}_{x'}$,
\begin{align*}
    \ell_\text{AE}(\rho_{x'}; {\theta}_\text{E}, {\theta}_\text{D}) 
    = 1 - \Tr[\rho_{x'}\Tilde{\rho}_{x'}].
\end{align*}
Together with the linear layer in the beginning, the QAE loss $\ell_\text{AE}$ is denoted as
\begin{align*}
        \ell_\text{AE}(\rho_{x'}; {\theta}_\text{E}, {\theta}_\text{D})
      = \ell_\text{AE}(x; \theta_\text{L}, {\theta}_\text{E}, {\theta}_\text{D}).
\end{align*}
We consider the ``trash'' qubits $N_{t}$ as a hyperparameter of QTTT.
Specifically, 
we discard $N_{t}$ in the output of the quantum encoding block and use the quantum decoder to reconstruct the quantum from the ``reference state''.
Further discussion on the necessity of trash qubit is provided in \cref{sec:ablation}.

\paragraph{Main Task Branch.} We select data re-uploading (\citet{perez2020data}) as the main task (classification) branch after the shared linear layer, amplitude encoding, and the variational encoding block. The main task block is denoted as a chain of products of parameterized unitaries,
\begin{align}\label{eq:data_reupload_VS}
    U_\text{M}(x'; {\theta}_\text{M}) = S_{L+1}({w}_{L+1}; x')\prod_{l=1}^L V_l({\phi}_{l}) S_l({w}_{l}; x'),
\end{align}
where ${\theta}_\text{M} = [{\phi}_{1}, {\phi}_{2}, \ldots, {\phi}_{L}, {w}_{1}, {w}_{2}, \ldots, {w}_{L+1}]$ is the parameter vector of the main task block. 
We use $S_l$ to denote the $l$-layer of data re-uploading layers and $V_l$ the $l$-layer of the variational layer.
The final state of the main task branch is denoted by $\rho_{\Tilde{y}}$,
\begin{align*}
    \rho_{\Tilde{y}}
    & = U_\text{M}(x'; {\theta}_\text{M})\rho_{z}U_\text{M}^\dagger(x'; {\theta}_\text{M})\\
    & = \rho_{\Tilde{y}}(x', \rho_z(x'; \theta_\text{E}); \theta_\text{M})\\
    & = \rho_{\Tilde{y}}(x'(x; \theta_\text{L}), \rho_z(x'(x; \theta_\text{L}); \theta_\text{E}); \theta_\text{M})\\
    & = \rho_{\Tilde{y}}(x; \theta_\text{L}, \theta_\text{E}, \theta_\text{M}).
\end{align*}
We measure each qubit individually and obtain the reduced density state $\rho_{\Tilde{y}, q}$. We introduce the relative fidelity between the single-qubit state $\rho_{\Tilde{y}, q}$ and the single-qubit label state $\rho_c$ of class $c$ as 
\begin{align*}
    f_{c,q}({x}; \theta_\text{L}, \theta_\text{E}, \theta_\text{M})
    = \Tr[{\rho}_c \rho_{\Tilde{y}, q}]\in[0, 1].
\end{align*}
We use $\hat{f}_{c, q}: d_x\times C\to [0, 1]$ to denote the objective fidelity regarding class $c$ given data pair $(x, y)$. The main task loss $\ell_\text{MT}$ is the weighted fidelity cost function of the final quantum state $\rho_y$ as 
\begin{align*}
\ell_\text{MT}(x, y; {\alpha}, {\theta}_\text{L}, {\theta}_\text{E}, {\theta}_\text{M}) 
= 
\frac{1}{2} \sum_{c=1}^{C}\sum_{q=1}^{N_q} \left[ \alpha_{c, q}\cdot f_{c, q}({x}; {\theta}_\text{L}, {\theta}_\text{E}, {\theta}_\text{M}) - \hat{f}_{c}(x, y) \right]^2,
\end{align*}
where ${\alpha}_{c, q}\in\R$ is the trainable \emph{class
weight} for qubit $q$ regarding class $c$ \citep{perez2020data}.

\begin{figure}[htb]
\centering
\begin{tabular}{ccc}
General Rotation Layer & \quad\quad\quad  & Ladder Entanglement Layer \\
\scalebox{0.7}{
\begin{quantikz}
        &      \setwiretype{q}&\gate[5]{\quad U_3\quad }&      &\midstick[5,brackets=none]{\quad=\quad}&\gate{U_3(\cdot, \cdot, \cdot)}\gategroup[5,steps=1,style={dashed,rounded corners, fill=gray_block, inner xsep=2pt},background,label style={label position=below,anchor=north,yshift=-0.2cm}]{} &\\
        &      \setwiretype{q}&                 &      &                             &\gate{U_3(\cdot, \cdot, \cdot)} &\\
        &      \setwiretype{q}&                 &      &                             &\gate{U_3(\cdot, \cdot, \cdot)} &\\
        &\vdots\setwiretype{n}&                 &\vdots&                             &\vdots                          &\\
        &      \setwiretype{q}&                 &      &                             &\gate{U_3(\cdot, \cdot, \cdot)} &
\end{quantikz}
}
& &
\scalebox{0.7}{
\begin{quantikz}
        &      \setwiretype{q}&\gate[5]{\quad E\quad }&      &\midstick[5,brackets=none]{\quad=\quad}
        &\ctrl{1}\gategroup[5, steps=4,style={dashed,rounded corners, fill=gray_block, inner xsep=2pt},background,label style={label position=below,anchor=north,yshift=-0.2cm}]{}
                                                               &\ghost{U_3}&\ \ldots\ &\targ{ }   &\\
        &      \setwiretype{q}&               &      &&\targ{ }&\ctrl{1}   &\ \ldots\ &\ghost{U_3}&\\
        &      \setwiretype{q}&               &      &&        &\targ{ }   &\ \ldots\ &\ghost{U_3}&\\
        &\vdots\setwiretype{n}&               &\vdots&&        &\vdots     &\ \ldots\ &\ghost{U_3}&\\
        &      \setwiretype{q}&               &      &&        &\ghost{U_3}&\ \ldots\ &\ctrl{-4}  &
\end{quantikz}
}
\end{tabular}
\caption{\textbf{General Rotation Layer (left) \& Ladder Entanglement Layer (right).}
The general rotation layer (left) consists of a series of parameterized general single qubit gates $U_3$ with three trainable parameters.
The entanglement layer (right) consists of a series of ladder CNOT gates for strong entanglement.
}
\label{fig:qc_bb}
\end{figure}
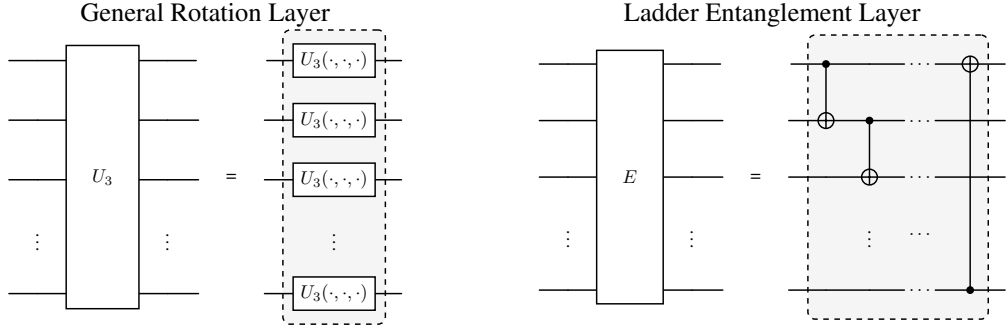

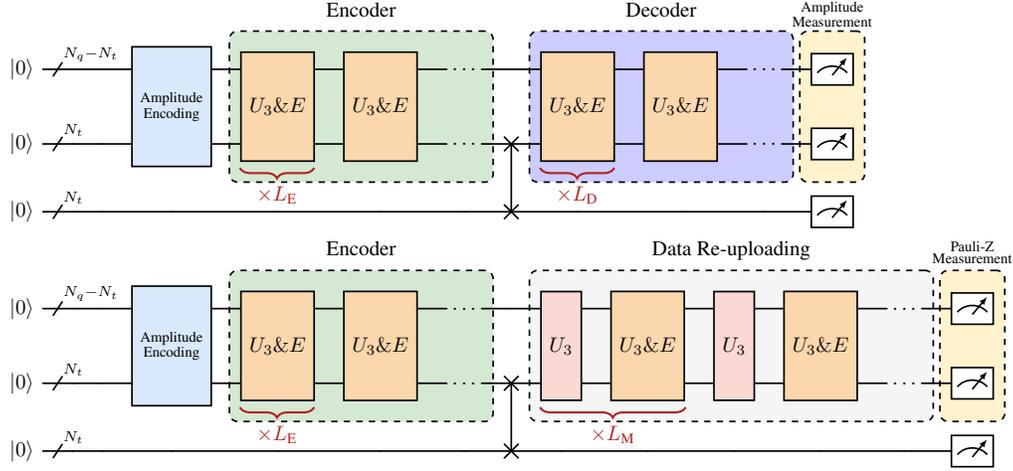
\begin{figure}[htb]
\centering
\begin{tabular}{l}
\scalebox{0.79}{
\begin{quantikz}%
        \lstick{$\ket{0}$} &\qwbundle{N_q-N_t}&
        &\gate[wires=2, style={fill=blue_ampEncoding},label style=black]{\substack{\text{Amplitude} \\ \text{Encoding}}}
        &\gate[wires=2, style={fill=orange_v}]{U_3\&E}\gategroup[2,steps=3,style={dashed,rounded corners, fill=green_encoder, inner xsep=2pt},background,label style={label position=above,anchor=north,yshift=+0.4cm}]{\text{Encoder}}
        &\gate[wires=2, style={fill=orange_v}]{U_3\&E}
        &\ \ldots \ & 
        &\gate[wires=2, style={fill=orange_v}]{U_3\&E}\gategroup[2,steps=3,style={dashed,rounded corners, fill=blue!20, inner xsep=2pt},background,label style={label position=above,anchor=north,yshift=+0.4cm}]{\text{Decoder}}
        &\gate[wires=2, style={fill=orange_v}]{U_3\&E}
        &\ \ldots \ & \meter{}\gategroup[2,steps=1,style={dashed,rounded corners, fill=yellow_measure, inner xsep=2pt},background,label style={label position=above,anchor=north,yshift=+0.4cm}]{$\substack{\text{Amplitude} \\ \text{Measurement}}$}\\
        \lstick{$\ket{0}$} &\qwbundle{N_t}     &&& &&\ \ldots \ &\swap{1}&& &\ \ldots \ &\meter{}\\
        \lstick{$\ket{0}$} &\qwbundle{N_t}     &&& &&           &\targX{}&& &           &\meter{}    
    \end{quantikz}
}
\\
\scalebox{0.79}{
\begin{quantikz}%
        \lstick{$\ket{0}$} &\qwbundle{N_q-N_t}&
        &\gate[wires=2, style={fill=blue_ampEncoding},label style=black]{\substack{\text{Amplitude} \\ \text{Encoding}}}
        &\gate[wires=2, style={fill=orange_v}]{U_3\&E}\gategroup[2,steps=3,style={dashed,rounded corners, fill=green_encoder, inner xsep=2pt},background,label style={label position=above,anchor=north,yshift=+0.4cm}]{\text{Encoder}}
        &\gate[wires=2, style={fill=orange_v}]{U_3\&E}
        &\ \ldots \ &        
        &\gate[wires=2, style={fill=pink_dru}]{U_3}\gategroup[2,steps=5,style={dashed,rounded corners, fill=gray_block, inner xsep=2pt},background,label style={label position=above,anchor=north,yshift=+0.4cm}]{\text{Data Re-uploading}}
        &\gate[wires=2, style={fill=orange_v}]{U_3\&E}
        &\gate[wires=2, style={fill=pink_dru}]{U_3}
        &\gate[wires=2, style={fill=orange_v}]{U_3\&E}
        &\ \ldots \ 
        & \meter{}\gategroup[2,steps=1,style={dashed,rounded corners, fill=yellow_measure, inner xsep=2pt},background,label style={label position=above,anchor=north,yshift=+0.4cm}]{$\substack{\text{Pauli-Z} \\ \text{Measurement}}$}\\
        \lstick{$\ket{0}$} &\qwbundle{N_t}    &&& &&\ \ldots \ &\swap{1}&&& &&\ \ldots \ &\meter{}\\
        \lstick{$\ket{0}$} &\qwbundle{N_t}    &&& &&           &\targX{}&&& &&           &\meter{}    
    \end{quantikz}
    \begin{tikzpicture}[overlay]
    \draw[line width=1pt, color=firebrick, decorate, decoration={brace, mirror, amplitude=5pt}] (-13.2,+3.6) -- (-11.95,+3.6) 
        node[midway, below=5pt] {\textcolor{firebrick}{$\times L_\text{E}$}};
    \draw[line width=1pt, color=firebrick, decorate, decoration={brace, mirror, amplitude=5pt}] (-8.15,+3.6) -- (-6.90,+3.6) 
        node[midway, below=5pt] {\textcolor{firebrick}{$\times L_\text{D}$}};
    \draw[line width=1pt, color=firebrick, decorate, decoration={brace, mirror, amplitude=5pt}] (-13.2,-0.43) -- (-11.95,-0.43) 
        node[midway, below=5pt] {\textcolor{firebrick}{$\times L_\text{E}$}};
    \draw[line width=1pt, color=firebrick, decorate, decoration={brace, mirror, amplitude=5pt}] (-8.15,-0.43) -- (-5.7,-0.43) 
        node[midway, below=5pt] {\textcolor{firebrick}{$\times L_\text{M}$}};
    \end{tikzpicture}
}
\end{tabular}
\caption{\textbf{Quantum Circuits of the QAE Branch (top) \& Main Task Branch (bottom).}
The QAE (top) consists of an amplitude encoding block, an encoder, a multi-quit SWAP operation, and a decoder. 
We make amplitude measurements at the end.
The main task branch (bottom) consists of the same amplitude encoding and the encoder in the QAE, followed by a data re-uploading block.
We make Pauli-Z measurements at the end.
The SWAP operations discard trash qubits by swapping the last two registers (both with $N_t$ qubits) of the quantum circuits.
The layer number in the encoder, decoder, and data re-uploading blocks are $L_{\text{E}}$, $L_{\text{D}}$, and $L_{\text{M}}$, respectively.
}
\label{fig:qc_2bs}
\end{figure}

During the main task inference (classification), we fix the weights $\alpha_{c, q} = \alpha_{c, q}^\star$ after training and select the class with the largest fidelity given data $x$,
\begin{align}
    \Tilde{y} = \argmax_{c}\sum_{q=1}^{N_q} \alpha_{c, q}^\star\cdot f_{c, q}({x}; {\theta}_\text{L}, {\theta}_\text{E}, {\theta}_\text{M}).
\end{align}
We use parameterized unitary $U_3(\cdot)$ with three Euler angles on each qubit and follow with strong entanglement (ladder CNOT) layers for both the encoder, decoder and the main QML model. 
We divide the input features by three and feed them into the $U_3$ gates.
On the other hand, QAE requires the qubit number of $\ceil{\log_2  d_x }$ for amplitude encoding.
Therefore, we choose the qubit number $\max ( \ceil{ d_x /3}, \ceil{\log_2  d_x } )$ and pad feature vectors with zeros to match the qubit configurations.
We run the experiment with state vector simulations. \cref{fig:qc_bb} shows the basic blocks and \cref{fig:qc_2bs} shows the quantum circuits of the two branches shown in \cref{fig:arch}.

\paragraph{Quantum Circuits.}
\cref{fig:qc_bb} presents the basic components of parameterized and entanglement blocks.
Specifically, we present the general rotation layer $U_3$ and ladder entanglement layer $E$.
Each $U_3$ layer has parameter numbers of 3 multiples of the qubit number, and each entanglement layer $E$ has the same number of CNOT gates of the qubit number.

\cref{fig:qc_2bs} shows the overall quantum circuit implementation of QTTT.
Note that each $U_3\&E$ layer represents a general rotation layer $U_3$ followed by a ladder entanglement layer $E$.
The QAE branch and the main task branch have the same number of qubits $N_q$, and $N_t$ is the number of trash qubits discarded after the encoder.
There are additional $N_t$ auxiliary qubits for resetting the qubit states after the quantum encoder.

Both the QAE branch and the main task branch are quantum circuits with total qubit number $N_q + N_t$.
The QAE branch (top of \cref{fig:qc_2bs}) consists of an encoding block and a decoding block. 
The first $N_q$ qubits encode the input data with amplitude encoding~\citep{mottonen2004transformation}. 
The non-negative trash qubits number $N_t$ is a hyper-parameter smaller than $N_q$.
The encoder is supposed to \emph{compress} the information into the first $N_q-N_t$ qubits and leave sparse information in the last $N_t$ qubits. 
We denote the number of $U_3\&E$ layers in the encoder as $L_\text{E}$.

The decoder takes both the quantum state of the first $N_q-N_t$ qubits out of the encoder and other $N_t$ qubits prepared in some reference states as input.
Between the encoder and decoder blocks, the SWAP operation realizes the above process by swapping the last two registers (both with $N_t$ qubits). 
Finally, we measure the amplitude of the final quantum state out of the decoder for later calculation of the QAE loss $\ssloss$.
We denote the number of $U_3\&E$ layers in the decoder as $L_\text{D}$.

The main task branch (bottom of \cref{fig:qc_2bs}) consists of the same encoder shared with the QAE followed by a data re-uploading block and Pauli-Z measurement at the end.
We apply the Pauli-Z measurement to calculate the main task loss $\mainloss$.
The alternatively repeated $V_l$ and the $S_l$ in \eqref{eq:data_reupload_VS} is realized with the $U_3$ and $U_3\&E$ gates, respectively.
We denote the repeat number of $U_3$ and $U_3\&E$ layers in the data re-uploading block as $L_\text{M}$. 

\paragraph{Multi-Task Loss.}
The total loss function $\ell$ is defined as the weighted sum of the main task loss $\ell_\text{MT}$ and the QAE loss $\ell_\text{AE}$:
\begin{align*}
    \ell(x, y; \sigma, {\alpha}, {\theta}_\text{L}, {\theta}_\text{E}, {\theta}_\text{D}, {\theta}_\text{M}) 
    = \frac{\ell_\text{MT}(x, y; {\alpha}, {\theta}_\text{L}, {\theta}_\text{E}, {\theta}_\text{M})}{2\sigma_{\text{MT}}^2} 
    + \frac{\ell_\text{AE}(x; {\theta}_\text{L}, {\theta}_\text{E}, {\theta}_\text{D})}{2\sigma_{\text{AE}}^2} + \log {(\sigma_{\text{MT}}\sigma_{\text{AE}})},
\end{align*}
where $\sigma=[\sigma_\text{MT}, \sigma_\text{AE}]$, and $\sigma_\text{MT}, \sigma_\text{AE} \in \R$ are trainable log-variance parameters for multi-task learning \citep{kendall2018multi}.
The purpose of $\sigma_\text{MT}$ and $\sigma_\text{AE}$ is weighting to balance between multi-task loss according to the noise level. 
Following \citet{kendall2018multi}, by maximizing the Gaussian likelihood, the model learns to scale the losses for each output.
A larger $\sigma$ represents a larger variance, thereby reducing the contribution of this task during optimization and vice versa.

\paragraph{Multi-Tasks Training.}
We minimize the following total loss on training dataset $D_\text{train}$ with number of data point $N_\text{test}$ while training:
\begin{align}\label{eqn:multi_task_loss}
    \sigma^\star, {\alpha}^\star, {\theta}_\text{L}^\star, {\theta}_\text{E}^\star, {\theta}_\text{D}^\star, {\theta}_\text{M}^\star 
    = 
    \argmin_{\sigma, {\alpha}, {\theta}_\text{L}, {\theta}_\text{E}, {\theta}_\text{D}, {\theta}_\text{M}}
    \frac{1}{N_\text{train}}\sum_{(x, y) \in D_\text{train}}\ell(x, y; \sigma, {\alpha}, {\theta}_\text{L}, {\theta}_\text{E}, {\theta}_\text{D}, {\theta}_\text{M}).
\end{align}

\paragraph{Test-Time Training.} After the regular training is completed, we fix the trained parameters ${\alpha}^\star$ and ${\theta}_\text{M}^\star$ in the main task branch. On the other hand, we initialize the encoder and decoder parameters with $({\theta}_\text{L}^\star,  {\theta}_\text{E}^\star, {\theta}_\text{D}^\star)$ and update them using gradient descent to $({\theta}_\text{L}^\circledast, {\theta}_\text{E}^\circledast, {\theta}_\text{D}^\circledast)$ to minimize the following QAE self-supervised loss: 
\begin{align}\label{eqn:ttt_loss}
    {\theta}_\text{L}^\circledast, {\theta}_\text{E}^\circledast, {\theta}_\text{D}^\circledast
    =
    \argmin_{{\theta}_\text{L}, {\theta}_\text{E}, {\theta}_\text{D}}\frac{1}{N_\text{test}}\sum_{(x, \cdot) \in D_\text{test}}
    \ell_\text{AE}(x; {\theta}_\text{L}, {\theta}_\text{E}, {\theta}_\text{D}).
\end{align}

By \cref{thm:main}, a one-step gradient descent on the QAE loss lowers the main task loss, we can deduce that minimizing the QAE loss \eqref{eqn:ttt_loss} during test-time training improves the performance of the model on the main task.

\section{Theoretical Analysis}\label{sec:theory}

In this section, we provide the theoretical analysis of our proposed methodology in \cref{sec:method}.
Specifically, in \cref{thm:main} we prove that under convex, smooth, and bounded loss functions, a one-step gradient during test time further lowers the main task loss.
In \cref{tab:complexity}, we analyze the computational complexity of QTTT to quantify the computation overhead ratio between QTTT test-time training and standard training. 

For clarity and fluency, we state some notations before stating the theorem. Because parameters $\theta_\text{L}$ and $\theta_\text{E}$ are parts of the encoding block the two branches share, we combine them and let $\theta_\text{LE} \coloneqq [\theta_\text{L}, \theta_\text{E}]$. Because in the following derivation, we focus on $\theta_\text{LE}$ and the gradient of the loss functions with respect to it, we neglect other irrelevant parameters without causing ambiguity.

\begin{theorem}\label{thm:main}
Let $\ell_\text{MT}(x, y; \theta_\text{LE})$ denote the main task loss on a test data point $(x, y)$, and $\ell_\text{AE}(x; \theta_\text{LE})$ the self-supervised auto-encoder task loss, which depends only on $x$. 
Here, $\theta_\text{LE}$ represents the shared parameters. 
Assume that for all $(x, y)$, main task loss $\ell_\text{MT}(x, y; \theta_\text{LE})$ is differentiable, convex, and $\beta$-smooth with respect to $\theta_\text{LE}$, and that both gradients are bounded.
\begin{align}
    {\|\nabla_{\theta_\text{LE}} \ell_\text{MT}(x, y; {\theta}_\text{LE})\|},  {\|\nabla_{\theta_\text{LE}} \ell_\text{AE}(x; {\theta}_\text{LE})\|} \leq G, \quad \forall \theta_\text{LE}.\label{eq:gradinets_bound}
\end{align}
With a fixed learning rate $\eta = \frac{\eps}{\beta G^2}$,  for every $(x, y)$ there exist $\epsilon > 0$ such that 
\begin{align}
\langle \nabla_{\theta_\text{LE}} \ell_\text{MT}(x, y; {\theta}_\text{LE}), \nabla_{\theta_\text{LE}} \ell_\text{AE}(x; {\theta}_\text{LE}) \rangle  > \epsilon.\label{eq:inner_product_lb}
\end{align}
Therefore, we have
\begin{align}
\ell_\text{MT}(x, y; {\theta}_\text{LE}^\prime)
<
\ell_\text{MT}(x, y; {\theta}_\text{LE}),
\end{align}
where $ \theta_\text{LE}^\prime =  \theta_\text{LE} - \eta \nabla_{\theta_\text{LE}} \ell_\text{AE}(x; {\theta}_\text{LE})$, i.e., Test-Time Training with one step of gradient descent.
\end{theorem}
\begin{remark}[Test-Time Training]
With a convex, smooth, and bounded loss function, we show that the main task loss on the testing dataset $D_\text{test}$ is further lowered when we first optimize all the parameters on the training data and then ``fine-tune'' the self-supervised QAE parameters on the testing data, that is
\begin{align*}
    \sum_{(x, y) \in D_\text{test}}\ell_\text{MT}\qty(x, y; {\alpha}^\star, {\theta}_\text{L}^\circledast, {\theta}_\text{E}^\circledast, {\theta}_\text{M}^\star) 
    < 
    \sum_{(x, y) \in D_\text{test}}\ell_\text{MT}\qty(x, y; {\alpha}^\star, {\theta}_\text{L}^\star, {\theta}_\text{E}^\star, {\theta}_\text{M}^\star).
\end{align*}
\end{remark}

\begin{proof}[Proof of \cref{thm:main}]
Our proof is based on \cite{sun2020test}. Given data point pair $(x, y)$, we expand the main task loss at $\theta_\text{LE}$ to the second order,
\begin{align}\label{eq:l_MT_expansion}
    \ell_\text{MT}(x, y;{\theta}_\text{LE} + \Delta{\theta}_\text{LE})
    \le \ell_\text{MT}(x, y;{\theta}_\text{LE}) 
    + \braket{\nabla_{\theta_\text{LE}}\ell_\text{MT}(x, y; {\theta}_\text{LE}), \Delta{\theta}_\text{LE}}
    + \frac{\beta}{2}\|\Delta{\theta}_\text{LE}\|^2.
\end{align}
Taking the step vector $\Delta{\theta}_\text{LE}$ and the step size $\eta^\star$:
\begin{align}
    \Delta{\theta}_\text{LE} 
    &= - \eta^\star \nabla_{\theta_\text{LE}}\ell_\text{AE}(x; \theta_\text{LE}),\label{eq:g_LE}\\
    \eta^\star 
    &= \frac{\braket{\nabla_{\theta_\text{LE}}\ell_\text{MT}(x, y; \theta_\text{LE}), \nabla_{\theta_\text{LE}}\ell_\text{AE}(x; \theta_\text{LE})}}{\beta\|\nabla_{\theta_\text{LE}}\ell_\text{AE}(x; \theta_\text{LE})\|^2}
    \ge\frac{\epsilon}{\beta G^2}.\label{eq:eta_star_lb}
\end{align}
We write \eqref{eq:l_MT_expansion} as
\begin{align}
       & \ell_\text{MT}(x, y;{\theta}_\text{LE} + \Delta{\theta}_\text{LE})\notag\\
      =& \ell_\text{MT}(x, y;{\theta}_\text{LE} - \eta^\star\nabla_{\theta_\text{LE}}\ell_\text{AE}(x; \theta_\text{LE}))\annot{By gradinet descent; \eqref{eq:g_LE}}\notag\\
    \le& \ell_\text{MT}(x, y;{\theta}_\text{LE}) 
    - \eta^\star\braket{\nabla_{\theta_\text{LE}}\ell_\text{MT}(x, y; {\theta}_\text{LE}),\ \nabla_{\theta_\text{LE}}\ell_\text{AE}(x; \theta_\text{LE})}
    + \frac{(\eta^\star)^2\beta}{2}\|\nabla_{\theta_\text{LE}}\ell_\text{AE}(x; \theta_\text{LE})\|^2\annot{By \eqref{eq:l_MT_expansion}}\notag\\
      =& \ell_\text{MT}(x, y;{\theta}_\text{LE}) 
    + \left(-\frac{1}{\beta} + \frac{1}{2\beta} \right)\cdot \frac{\|\braket{\nabla_{\theta_\text{LE}}\ell_\text{MT}(x, y; {\theta}_\text{LE}),\ \nabla_{\theta_\text{LE}}\ell_\text{AE}(x; \theta_\text{LE})}\|^2}{\|\nabla_{\theta_\text{LE}}\ell_\text{AE}(x; \theta_\text{LE})\|^2}\annot{By \eqref{eq:eta_star_lb}}\notag\\
    \le& \ell_\text{MT}(x, y;{\theta}_\text{LE}) 
    -\frac{\epsilon^2}{2\beta G^2}\label{eq:loss_difference_square},
\end{align}
where the last inequality \eqref{eq:loss_difference_square} is from \eqref{eq:gradinets_bound} and \eqref{eq:inner_product_lb}. 
Rearranging \eqref{eq:loss_difference_square}, we have
\begin{align}
    \ell_\text{MT}(x, y;{\theta}_\text{LE})
    - 
    \ell_\text{MT}(x, y;{\theta}_\text{LE} - \eta^\star\nabla_{\theta_\text{LE}}\ell_\text{AE}(x; \theta_\text{LE}))
    \ge
    \frac{\epsilon^2}{2\beta G^2}.\label{eq:loss_difference_square_rearranged}
\end{align}
Let $\eta\in(0, \eta^\star]$. By convexity of $\ell_\text{AE}$ we have
\begin{align}
   & \ell_\text{MT}(x,y; \theta_\text{LE} + \Delta\theta_\text{LE})
  =  \ell_\text{MT}(x,y; \theta_\text{LE} -\eta\nabla_{\theta_\text{LE}}\ell_\text{AE}(x; \theta_\text{LE}))\notag\\
  =& \ell_\text{MT}\left({x,y; \left({1 - \frac{\eta}{\eta^\star}}\right) \theta_\text{LE} 
   + \frac{\eta}{\eta^\star} {( \theta_\text{LE} - \eta^\star\nabla_{\theta_\text{LE}}\ell_\text{AE}(x; \theta_\text{LE}) )}}\right)\notag\\
\le& {\left(1 - \frac{\eta}{\eta^\star}\right)} \ell_\text{MT}(x,y; \theta_\text{LE}) 
   + \frac{\eta}{\eta^\star} \ell_\text{MT}(x,y; \theta_\text{LE} - \eta^\star\nabla_{\theta_\text{LE}}\ell_\text{AE}(x; \theta_\text{LE})).\notag
\end{align}
Combining with \eqref{eq:loss_difference_square_rearranged}, we have
\begin{align*}
 \ell_\text{MT}(x,y; \theta_\text{LE} -\eta\nabla_{\theta_\text{LE}}\ell_\text{AE}(x; \theta_\text{LE}))
&\leq {\left(1 - \frac{\eta}{\eta^\star}\right)} \ell_\text{MT}(x,y; \theta_\text{LE})
+ \frac{\eta}{\eta^\star} {\left(\ell_\text{MT}(x,y; \theta_\text{LE}) - \frac{\epsilon^2}{2\beta G^2}\right)} \\
&= \ell_\text{MT}(x,y; \theta_\text{LE}) - \frac{\eta}{\eta^\star}\frac{\eps^2}{2\beta G^2}.
\end{align*}
Since $\eta / \eta^\star > 0$, we have
\begin{align}
\ell_\text{MT}(x, y;{\theta}_\text{LE} - \eta\nabla_{\theta_\text{LE}}\ell_\text{AE}(x; \theta_\text{LE}))
<
\ell_\text{MT}(x, y;{\theta}_\text{LE}).
\end{align}
This completes the proof.
\end{proof}

\paragraph{Computational Complexity.}
We summarize the computational complexity for each component of QTTT in \cref{tab:complexity}.
We calculate the gradient of the PQC with the parameter-shift rule.
We treat the number of gate operations as the basic unit, assuming all gates operate individually and separately, and neglect the classical computation before and after the quantum circuits. Here, we use the same notation as introduced in the main paper:
$N_\text{train}$ represent the number of training data,
$N_\text{test}$ the number of testing data,
$N_q$ the number of qubits,
$L_\text{E}$ the number of layers of the encoder, 
$L_\text{D}$ the number of layers of the decoder, 
$L_\text{M}$ the number of layers of the data re-uploading block in the main task branch, 
and
$S$ is the total number of measurement shots.
From \cref{tab:complexity}, the proportion of training gate complexity (time complexity) for computing gradient with TTT is $\Theta \left( {N_{\text{test}}}/{N_{\text{train}}}\times \left({\layerE + \layerD} \right)^2 / \left({\layerE + \layerD + \layerM} \right) ^2 \right)$.
For example, assuming $\layerE \approx \layerD \ll \layerM $ the overhead for QTTT is of the ratio $\Theta \left( \Ntest/(\Ntrain \times\layerM)\right)$, with the testing data generally being a small portion compared to the training data.
Notably, the overhead ratio of TTT scales is inversely proportional to the main task model depth $\layerM$, meaning that the deeper the models, the smaller the overhead ratio, making QTTT an ideal plug-and-play framework for other QML models.

\begin{table*}[htp!]
\centering
\vspace{-0.5em}
    \begin{tabular}{cl}
    \toprule
    Aspect & Gate Complexity \\
    \midrule
        Gradient Calculation (Training) & $\Theta \left( S \times N_{\text{train}} \times N_q^2 \times \left( L_{\text{E}}+L_{\text{D}}+L_{\text{M}} \right)^2 \right)$  \\
        Gradient Calculation (QTTT) & $\Theta \left( S \times N_{\text{test}} \times N_q^2 \times (L_{\text{E}} + L_{\text{D}})^2 \right)$  \\
    \bottomrule
    \end{tabular}
     \label{tab:complexity}
\caption{
\textbf{Computational Complexity.}
We summarize the gate complexities for the gradient calculation of the PQC with the parameter-shift rule.
}
\end{table*}

\section{Experiments}\label{sec:exp}

In this section, we demonstrate QTTT's ability to (1) tackle distribution shifts in the testing data and (2) mitigate random unitary noise in quantum circuits during inference.
We select four QML models as the baseline from the top performance models reported by the benchmark paper \citep{bowles2024better}.
It consists of two variational QML models: data reuploading classifier \citep{perez2020data} and dressed quantum circuit \citep{mari2020transfer}; two kernel-based QML models: IQP kernel classifier \citep{havlivcek2019supervised} and projected quantum kernel \citep{huang2021power}.
In our experimental setup, we set the quantum circuit layers $L_{\text{E}}=L_{\text{D}}=L_{\text{M}}=4$ (see \cref{fig:qc_2bs}).

Our experiments are organized into the following three parts, respectively showing:
\begin{itemize}
    \item QTTT demonstrates robustness against data corruption, including Gaussian noise, brightness variations, fog, and snow effects, outperforming the state-of-the-art QML model (see \cref{sec:dis_shift}).
    \item QTTT mitigates simple random unitary noise in quantum circuits at test time, paving the way for a new noise-aware architecture design in QML models as well as a plug-and-play extension for existing QML models (see \cref{sec:noisy_qc}).
    \item We conduct ablation studies to verify the effectiveness of QTTT architectural design (see \cref{sec:ablation}).
\end{itemize}

\subsection{Distribution Shift}\label{sec:dis_shift}
We follow the data corruption settings from \citet{gandelsman2022test,sun2020test} and implement the baseline models and datasets from the benchmark paper \citep{bowles2024better}. 
For details on the dataset generation process, see \citet{bowles2024better}; the datasets consist of $N_d=300$ samples with a training/testing split of $4:1$.

\paragraph{Setup.} We select top-performance QML models based on the benchmark results in \citet{bowles2024better}.
We follow the model training settings for QTTT and the baseline QML models.
These five synthetic datasets include linear separable, hidden manifold, two curves, hyperplanes, bar \& stripes \citep{bowles2024better}.
We exclude the MNIST data since it either needs to drastically downsize the image or perform PCA to reduce feature dimensions.
We average the results over the five kinds of datasets with three random seeds for each, hence 15 experiments for each.
We select the datasets with feature dimensions $d_x=5$ and $d_x=10$, and the corresponding qubits number is $N_q=3$ and $N_q=4$, respectively. 
For both experiments, we choose $N_t=0$.
Here, we summarize the four corruptions we employ in this section.
\begin{itemize}[leftmargin=2.4em]
\item [\textbf{(C1)}]\label{item:noise2} \textbf{Brightness Change.} 
We change the brightness of the data by perturbing the testing data by a factor, i.e., $x_\text{perturbed} = x \times (\text{Factor})$.
\item [\textbf{(C2)}]\label{item:noise3} \textbf{Fog Effect.} 
We apply the fog effect on data by perturbed testing data, $ x_\text{perturbed} = (1-\text{Intensity})\times x + (\text{Intensity}) \times \text{Mean}(x)$.
\item [\textbf{(C3)}]\label{item:noise4} \textbf{Snow Effect.} 
We apply the snow effect on data by adding random noise (uniform distribution) to each feature with probability $p$.
\item [\textbf{(C4)}]\label{item:noise1} \textbf{Gaussian Noise.} 
We add Gaussian noise to the testing sets by perturbing the testing data $D_\text{test}=\{(x, \cdot)_i\}$ with noise sampled from the normal distribution.
Specifically, the noise added to each element of $x$, the final pertubed data is computed as $x_\text{perturbed} = x + \text{(noise level)} \times \calN(0, 1)$.
\end{itemize}

\begin{figure}[htb!]
    \centering
    \text{$d_x=5,\ N_q=3,\ N_t=0$}\\
    \begin{minipage}{\textwidth}
        \centering
        \hspace{-1.0em}
        \subfloat[Brightness Change.]{%
            \includegraphics[width=0.25\textwidth]{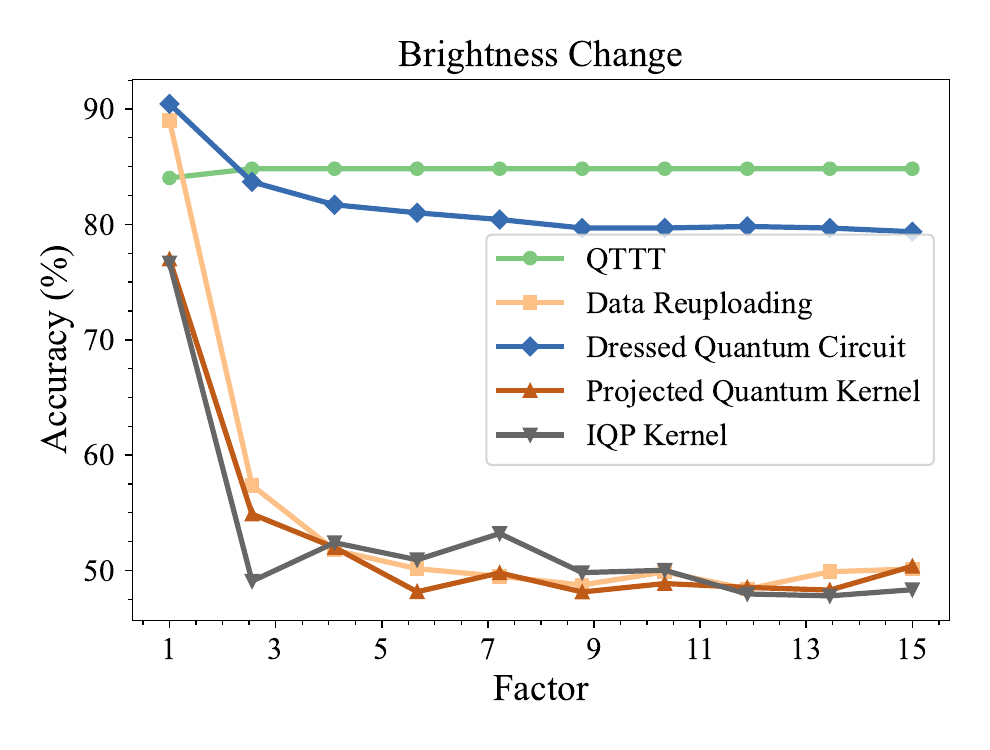}
            \label{fig:brightness}
        }
        \hspace{-1.0em}
        \subfloat[Fog Effect.]{%
            \includegraphics[width=0.25\textwidth]{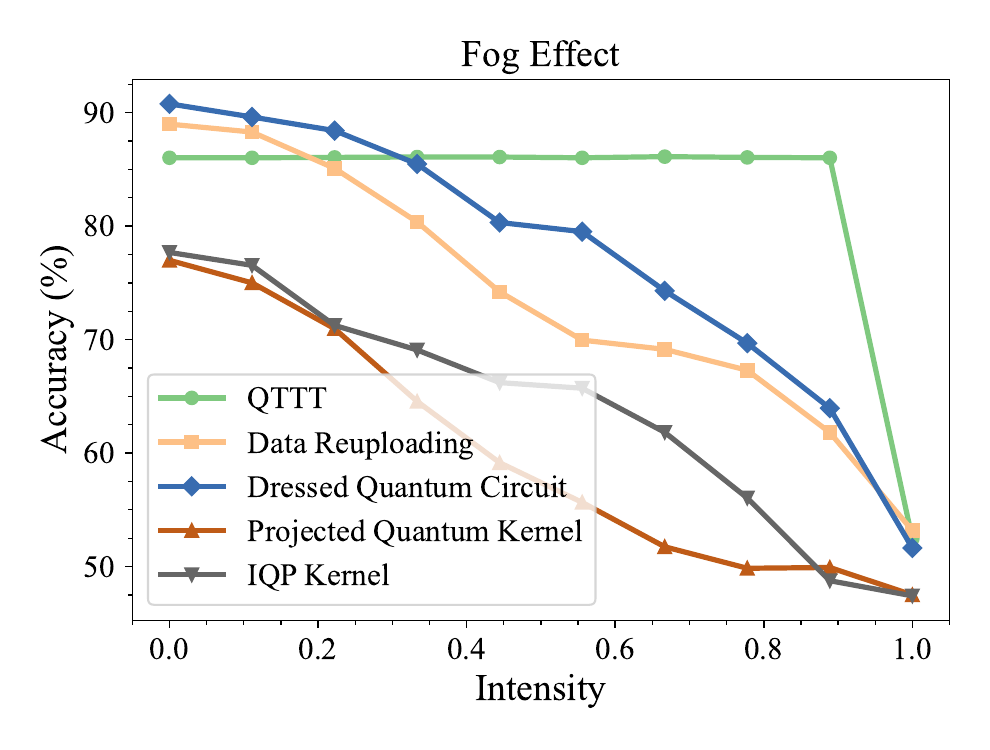}
            \label{fig:fog}
        }
        \hspace{-1.0em}
        \subfloat[Snow Effect.]{%
            \includegraphics[width=0.25\textwidth]{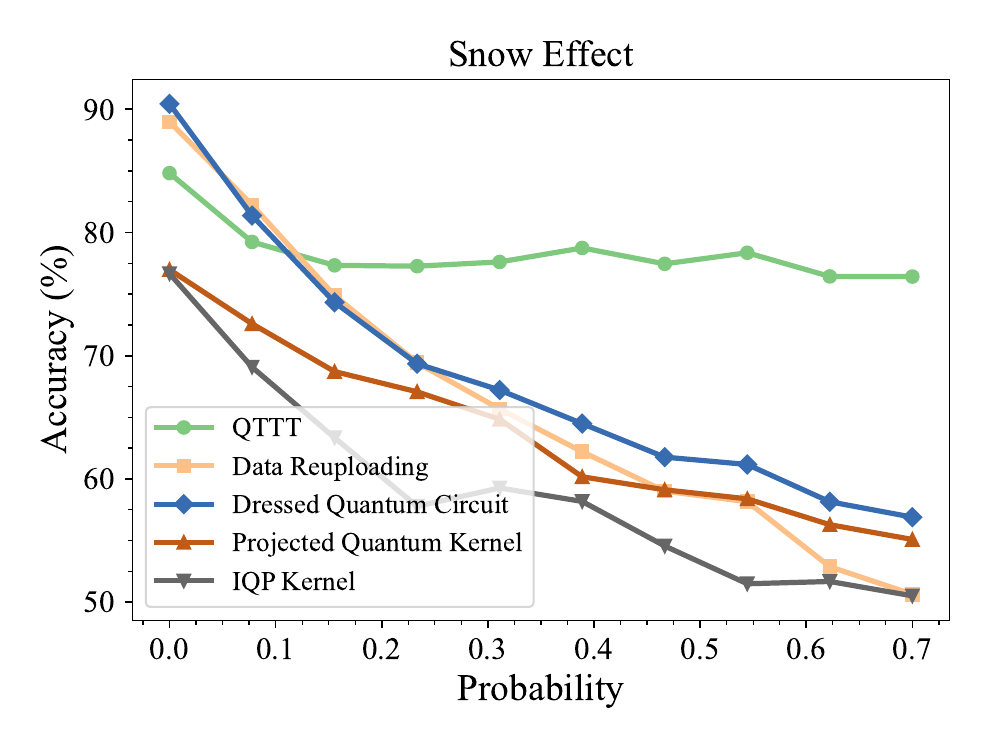}
            \label{fig:snow}
        }
        \subfloat[Gaussian Noise.]{%
            \includegraphics[width=0.25\textwidth]{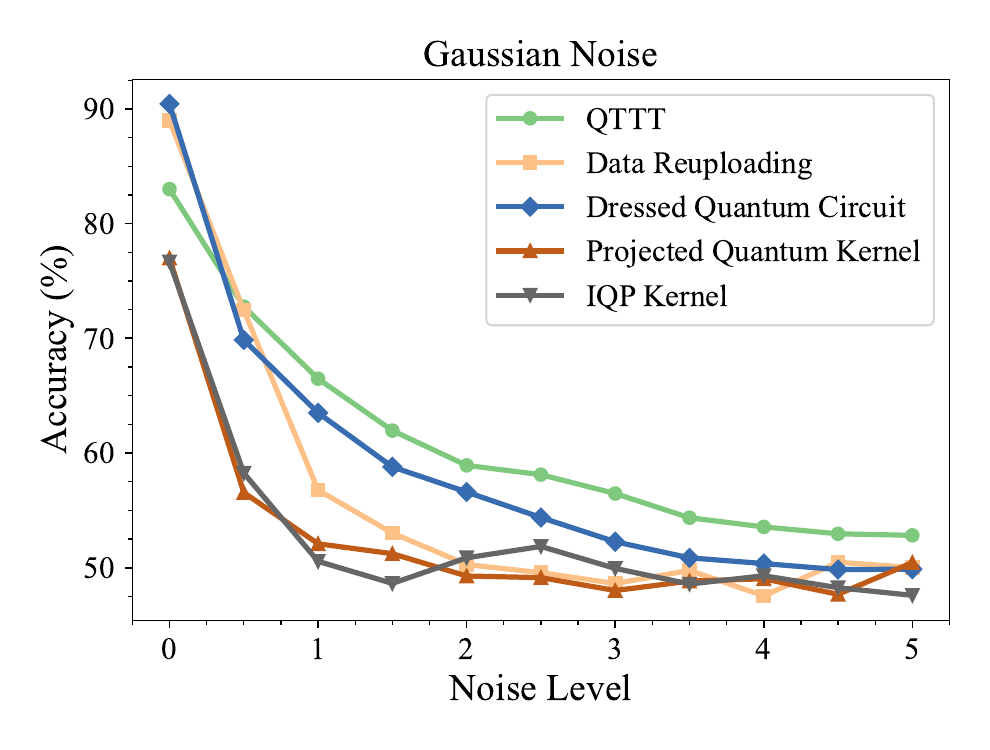}
            \label{fig:gauss}
        }
    \end{minipage}
    
    \vspace{0.4cm} 
    
    \text{$d_x=10,\ N_q=4,\ N_t=0$}\\
    \begin{minipage}{\textwidth}
        \centering
        \hspace{-1.0em}
        \subfloat[Brightness Change.]{%
            \includegraphics[width=0.25\textwidth]{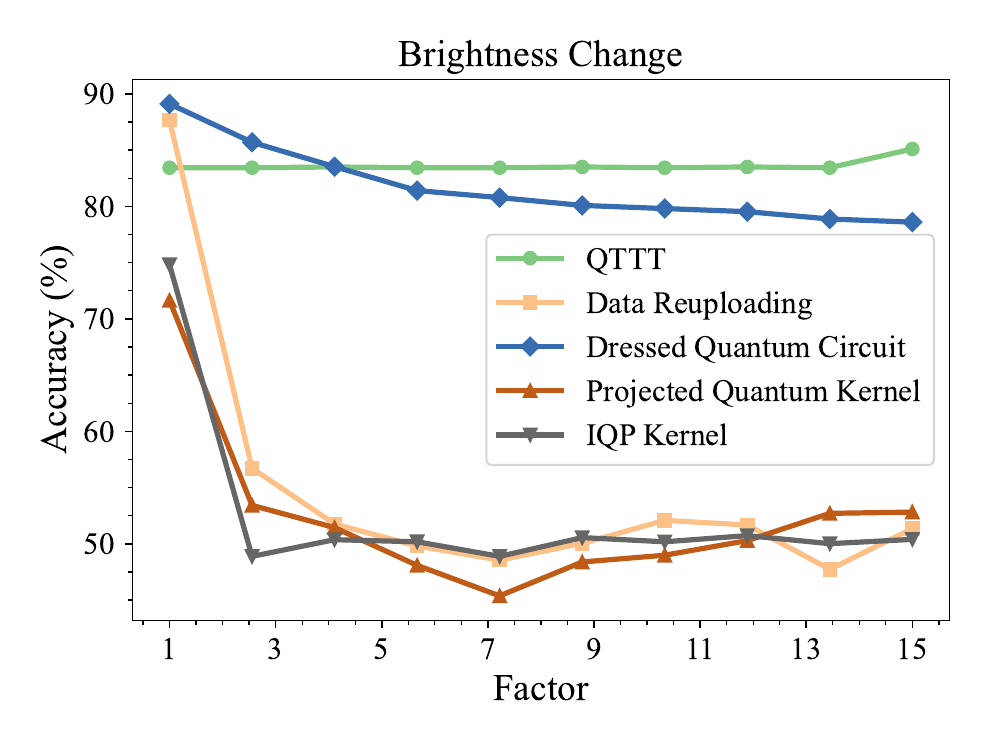}
            \label{fig:brightness10}
        }
        \hspace{-1.0em}
        \subfloat[Fog Effect.]{%
            \includegraphics[width=0.25\textwidth]{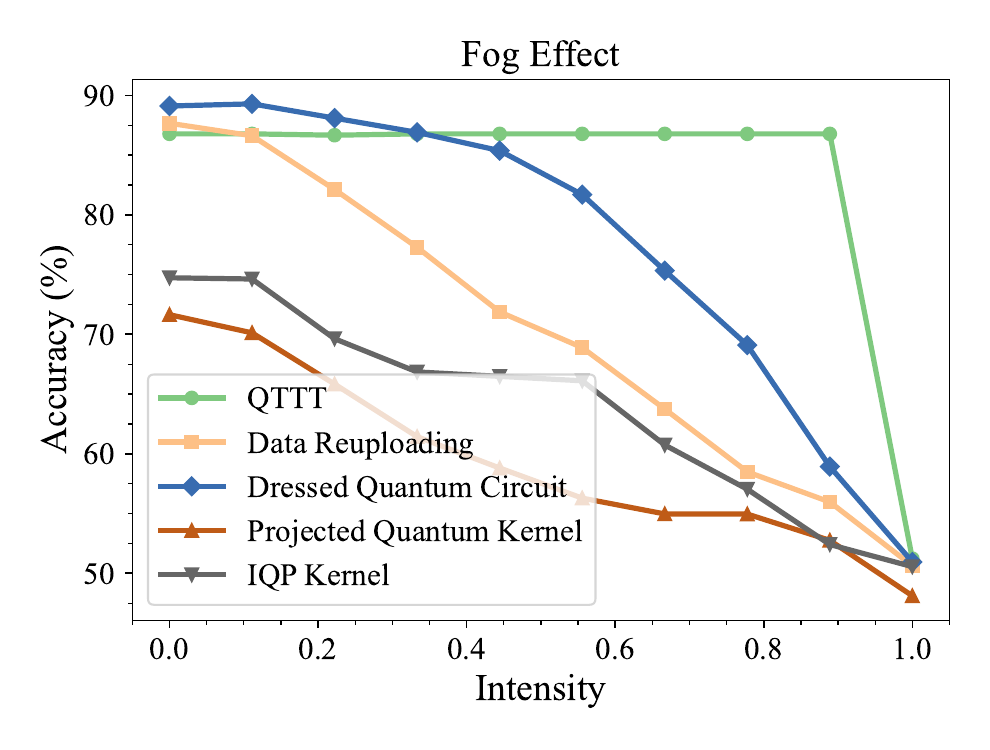}
            \label{fig:fog10}
        }
        \hspace{-1.0em}
        \subfloat[Snow Effect.]{%
            \includegraphics[width=0.25\textwidth]{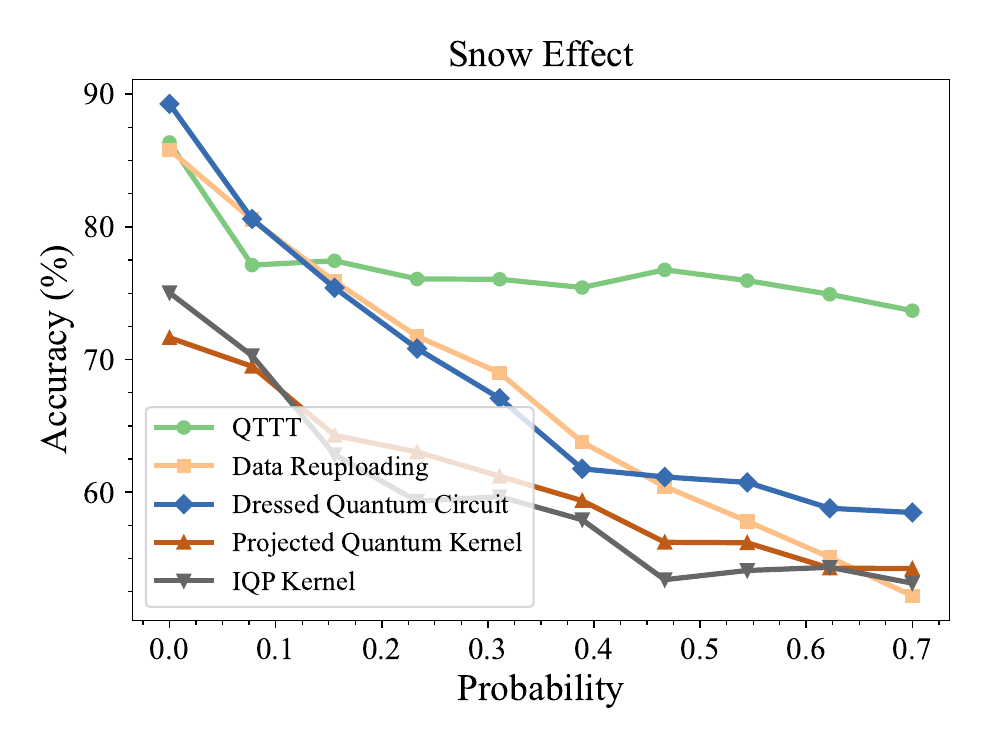}
            \label{fig:snow10}
        }
        \subfloat[Gaussian Noise.]{%
            \includegraphics[width=0.25\textwidth]{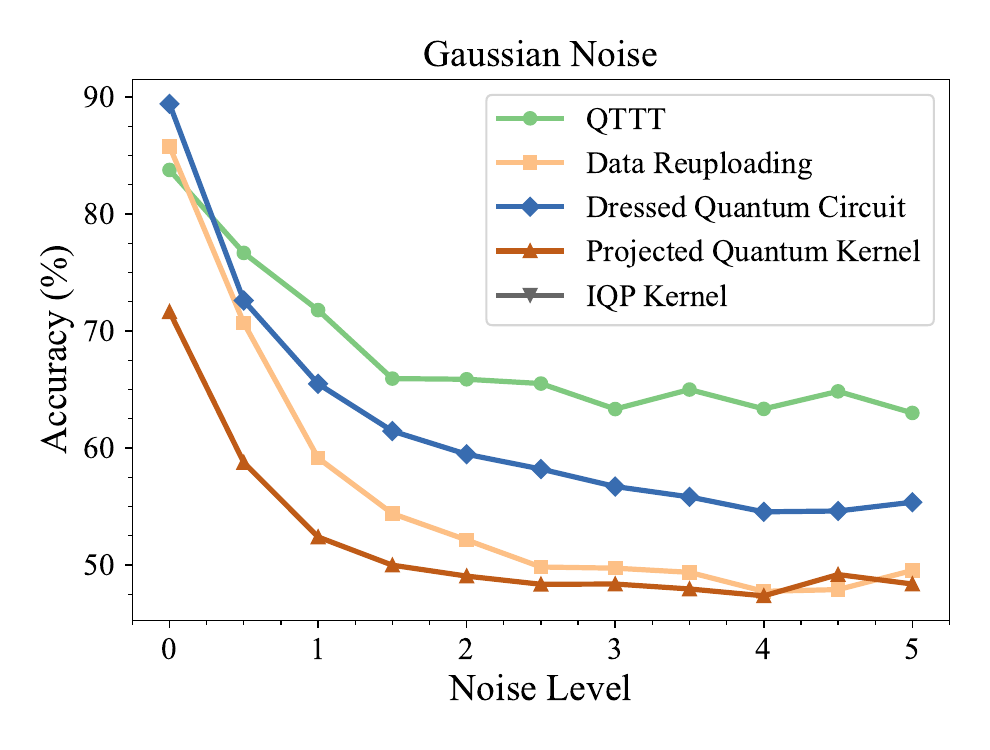}
            \label{fig:gauss10}
        }
    \end{minipage}
    
    \caption{\textbf{Test Accuracy (\%) on QML-Benchmark with Different Levels of Noise.}
    Data corruptions are applied to the testing datasets.
    We use the dataset with feature dimensions $d_x=5$ and $d_x=10$ for the experiments of the top and bottom rows, respectively. 
    The results indicate that QTTT maintains robustness as noise levels increase, whereas the performance of other models declines more rapidly. 
    Moreover, the performance gap between QTTT and other QML models widens as data dimensions increase.
    }
    \label{fig:dist_shift}
\end{figure}

\paragraph{Results.}
In \cref{fig:dist_shift}, we demonstrate the model accuracy on the testing datasets under different data corruptions.
For \cref{fig:brightness} to \cref{fig:gauss} we use the dataset with feature dimension $d_x=5$.
For \cref{fig:brightness10} to \cref{fig:gauss10} we use $d_x=10$.
For \cref{fig:gauss}, the accuracy consistently outperforms all baseline QML methods when adding Gaussian noise.
For \cref{fig:brightness,fig:fog,fig:snow}, QTTT exhibits more robustness against increasing corruption levels.
Remarkably, with increasingly corrupted data, QTTT still maintains its performance while others drop off rapidly.
However, the QTTT performance on original uncorrupted data is slightly lower than that of the data reuploading classifier and dressed quantum circuit.
The slight performance drop is due to the multi-task learning from two objectives: minimizing the self-supervised loss and minimizing the main classification task loss.

\subsection{Noisy Quantum Circuit}\label{sec:noisy_qc}

In a real-world application scenario, the QML model is trained on a noise-free quantum computer, while the inference can be performed on a noisy quantum computer. 
The QML model needs to be robust against quantum circuit errors. 
In this section, we demonstrate QTTT's ability to mitigate a simple noise model, specifically random unitary noise. 

\paragraph{Setup.}
After each layer, we apply a rotation gate randomly sampled from $\{R_x, R_y, R_z\}$ with equal probability $p=1/3$ on each qubit, with each rotation angle sampled from $\text{Unif}(0, \epsilon_\theta)$. 
We use data re-uploading as the baseline for this experiment and follow the same experimental setup and training setting as described in \cref{sec:dis_shift}. 
We choose $d_x=5,\ N_q = 3,\ N_t=0$ for the experiment.
We still average the results over the five kinds of datasets $\times$ ten random seeds, hence 50 experiments for each noise level.
Note that all models are trained in a noise-free setting. However, the random unitary noise described above is applied during inference. 
We examine both batch and online versions of QTTT, where data is provided either in batches or sequentially.

\begin{figure}[htp!]
    \centering
    \begin{minipage}[t]{0.48\linewidth}
        \centering
        \includegraphics[width=0.96\linewidth]{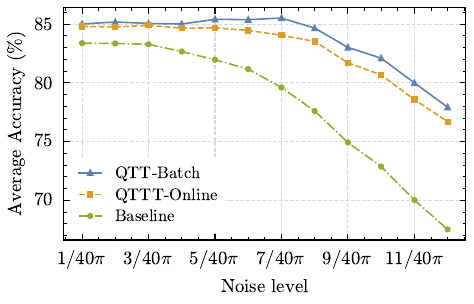}
    \end{minipage}
    \hspace{-0.5em}
    \begin{minipage}[t]{0.48\linewidth}
        \centering
        \includegraphics[width=\linewidth]{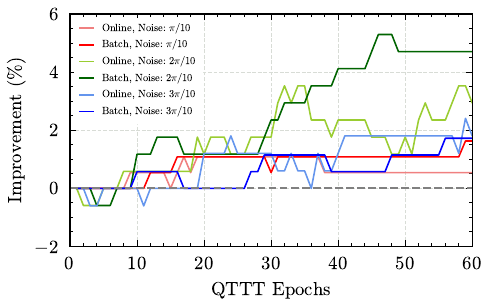}
    \end{minipage}
        \caption{
        \textbf{Test Accuracy and Improvements (\%) for Noisy Quantum Circuits.}
        (left) We report the average test accuracy with QTTT across all datasets, with each dataset generated three times using different random seeds under increasing noise levels. 
        (right) We report test accuracy improvements under three noise levels, averaging over three random seeds for each point. The darker colors indicate the batch version of QTTT, while lighter colors represent the online version. 
        In every case (see detailed results in \cref{tab:noisy_circuit}), the batch version of QTTT slightly outperforms the online version.
        }
        \label{fig:noisy_ttt}
\end{figure}

\paragraph{Results.}
In \cref{tab:noisy_circuit}, we summarize the performance of QTTT and the baseline models in noisy quantum circuits under increasing circuit noise levels. 
QTTT with batch optimization during test time achieves a $7.0\%$ improvement on average in noisy circuit settings. 
Meanwhile, the online version of QTTT still achieves an improvement of $5.9\%$, despite only ``overfitting'' on a single test sample.
QTTT offers a plug-and-play extension for existing QML models when inference on a noisy quantum circuit.
We demonstrate that QTTT mitigates the simple random unitary noise and improves the performance by up to $7.0\%$.
In \cref{fig:noisy_ttt}, we visualize the average accuracy of increasing noise levels and the improvement of the model accuracy with three noise levels over test-time training epochs. 

\begin{table*}[htb!]
\centering
\resizebox{ \textwidth}{!}{%
\begin{tabular}{lccccccccccccccc}
\toprule
$\text{Unif}(0,\epsilon_\theta)$ & $\pi/40$ & $2\pi/40$ & $3\pi/40$ & $4\pi/40$ & $5\pi/40$ & $6\pi/40$ & $7\pi/40$ & $8\pi/40$  & $9\pi/40$  & $10\pi/40$ & $11\pi/40$ & $12\pi/40$ & Mean \\
\midrule
Data Re-uploading & 83.374 & 83.351 & 83.281 & 82.665 & 81.967 & 81.167 & 79.604 & 77.606 & 74.927 & 72.879 & 70.036 & 67.534 & 78.199 \\
QTTT-Online & 84.802 & 84.756 &  84.893 &  84.653 & 84.678 &  84.463 &  84.052 & 83.525 & 81.702 & 80.675 &  78.579 & 76.698 & 82.790 \\
\rowcolor{almond} QTTT-Batch (Default) & \textbf{85.007} & \textbf{85.178} &  \textbf{85.053} &  \textbf{85.001} & \textbf{85.404} & \textbf{85.378} & \textbf{85.511} & \textbf{84.656} & \textbf{83.016} & \textbf{82.095} & \textbf{79.996} & \textbf{77.93} & \textbf{83.685} \\
\bottomrule
\end{tabular}
}
\vspace{-0.5em}
\caption{
\textbf{
QTTT on Noisy Quantum Circuit.}
We select twelve noise level upper bounds $\epsilon_\theta$ from $\pi/40$ to $12\pi/40$, applying random rotation gates with angle sampled from $\text{Unif}(0,\epsilon_\theta)$ after each layer of the QML models.
QTTT-Online and QTTT-Batch improve the model accuracy by $7.0\%$ and $5.9\%$, respectively, compared to the vanilla data re-uploading model.
}
\label{tab:noisy_circuit}
\end{table*}

\subsection{Ablation Studies}\label{sec:ablation}
We compare the model accuracy for batch and online learning and the trash qubit number of the QAE (see details in \cref{sec:method}). 
We also conduct comprehensive ablation studies on QTTT and validate the QTTT configuration setups.
We follow the same experimental setup and training setting described in \cref{sec:dis_shift} and choose $d_x=5,\ N_q = 3$.
The results are summarized in \cref{tab:ablation}, where the top half compares various QTTT configurations such as batch learning, online learning, trash qubit number $N_{t}=1$ and $N_{t}=2$, and for the bottom half we remove various components of QTTT as ablation study, including removing test-time training, the linear layer, data reuploading in the main task branch, and the multi-task loss (see \cref{sec:method}).
We average the model accuracy over the $5 (\text{datasets}) \times 3(\text{seeds}) =15$ experiments.

The results in \cref{tab:ablation} indicate that either the default or a moderate number of trash qubits $N_{t}$ yields better model performance across various corruptions and quantum circuit errors. 
Additionally, we verified the effectiveness of our architectural design by examining various components of QTTT, including test-time training, the linear layer, data reuploading in the main task branch, and multi-task objective training.

\begin{table*}[htb!]
\centering
\resizebox{\textwidth}{!}{%
\begin{tabular}{l|ccccccc|c}
\toprule
\multicolumn{1}{l|}{} & Brightness & Fog & Snow & Gaussian  & $\pi/10$ & $2\pi/10$ & $3\pi/10$ & Mean \\
\midrule
\rowcolor{almond} \multicolumn{1}{l|}{QTTT-Batch (Default)}            & \underline{86.88} & \underline{86.94} & \textbf{73.23} & 52.59  & 85.00 & \textbf{84.66} & \textbf{77.93} & \textbf{78.18} \\
\multicolumn{1}{l|}{QTTT w/ $N_{t}=1$}   & 87.40 & \textbf{87.40} & 72.72 & \textbf{54.40} & 83.80 & 73.42 & 65.88 & 75.00 \\
\multicolumn{1}{l|}{QTTT w/ $N_{t}=2$}         & \textbf{87.83} & 81.67 & 71.64 & \underline{53.39}  & \underline{86.73} & \underline{76.96} & \underline{67.43} & \underline{75.09} \\
\multicolumn{1}{l|}{QTTT-Online}          & 47.45 & 47.47 & 43.44 & 30.52  &  86.72 & 72.17 & 66.70 & 56.35 \\
\midrule
\multicolumn{1}{l|}{w/o TTT}                 & 85.73 & 75.83 & 58.31 & 50.40 & 83.38 & 62.93 & 56.69 & 67.61 \\
\multicolumn{1}{l|}{w/o Linear Layer}        & 84.20 & 84.20 & 71.26 & 49.93 & 74.58 & 59.59 & 52.90 & 68.09 \\
\multicolumn{1}{l|}{w/o Data Reuploading}    & 63.63 & 63.63 & 62.33 & 49.07  & 66.58 & 59.24 & 56.02 & 60.07 \\
\multicolumn{1}{l|}{w/o Multi-Task Loss}    & 86.70 & 86.73 & \underline{72.82} & 52.40  & \textbf{87.81} & 73.21 & 62.93 & 74.66 \\
\bottomrule
\end{tabular}%
}
\vspace{-0.5em}
\caption{
\textbf{
Component Ablation Studies of QTTT.}
The best and the second best accuracy across different corruption types and quantum circuit noise level upper bounds are shown in \textbf{bold} and with \underline{underlining}, respectively.
}
\label{tab:ablation} %
\end{table*}

\section{Limitations and Future Works.}\label{sec:lim}

To facilitate the real-world application of QML models, future studies can explore more sophisticated QTTT architectures and other self-supervised tasks beyond quantum auto-encoders by quantum state recovery.
Moreover, the current test-time training optimization relies on first-order gradient descent. 
Employing higher-order or quantum-aware optimizers~\citep{huang2024l2o,stokes2020quantum,gacon2021simultaneous} could enhance its performance.
While we experiment with QTTT under various data corruptions and quantum circuit noise levels, benchmarking with more advanced noise models will further support the real-world deployment of QTTT on actual quantum hardware.
Additionally, a theoretical understanding of the trainability of QTTT and its limitations in a noisy setting would be valuable.

\section{Conclusion}\label{sec:conclusion}
In this paper, we introduce QTTT, a novel QML framework designed for future real-world applications. QTTT serves as a plug-and-play extension for existing QML models. 
QTTT demonstrates robustness to distribution shifts between training and testing data, as well as mitigating simple quantum circuit noise during inference. 
This is achieved by minimizing the self-supervised loss of a quantum auto-encoder. 
Empirically, we show that QTTT outperforms existing QML models on corrupted data and random unitary noise introduced at test time. 
Theoretically, we provide performance guarantees for QTTT and analyze its computational complexity, showing its minor overhead with increased quantum model depth.

However, scaling up QML in the NISQ era poses challenges in demonstrating whether it offers advantages over classical counterparts. 
These challenges include the expressiveness and trainability of QML models, as well as error mitigation on real quantum hardware.
Our work takes a small step toward unlocking real-world applications of QML in the NISQ era, providing a novel noise-aware framework for existing QML models.

\clearpage
\section*{Acknowledgements}
H.-S.G. acknowledges support from the National Science and Technology Council, Taiwan under Grants No. NSTC 113-2112-M-002-022-MY3, No. NSTC 113-2119-M-002 -021, No. NSTC112-2119-M-002-014, No. NSTC 111-2119-M-002-007, and No. NSTC 111-2627-M-002-001, from the US Air Force Office of Scientific Research under Award Number FA2386-23-1-4052 and from the National Taiwan University under Grants No. NTU-CC-112L893404 and No. NTU-CC-113L891604. H.-S.G. is also grateful for the support from the “Center for Advanced Computing and Imaging in Biomedicine (NTU-113L900702)” through The Featured Areas Research Center Program within the framework of the Higher Education Sprout Project by the Ministry of Education (MOE), Taiwan, and the support from the Physics Division, National Center for Theoretical Sciences, Taiwan.

\def\arxivfont{\rm}
\bibliography{reference}

\begin{thebibliography}{}

\bibitem[Abbas et~al., 2021]{abbas2021power}
Abbas, A., Sutter, D., Zoufal, C., Lucchi, A., Figalli, A., and Woerner, S. (2021).
\newblock The power of quantum neural networks.
\newblock {\em Nature Computational Science}, 1(6):403--409.

\bibitem[Amin et~al., 2018]{amin2018quantum}
Amin, M.~H., Andriyash, E., Rolfe, J., Kulchytskyy, B., and Melko, R. (2018).
\newblock Quantum boltzmann machine.
\newblock {\em Physical Review X}, 8(2):021050.

\bibitem[Bowles et~al., 2024]{bowles2024better}
Bowles, J., Ahmed, S., and Schuld, M. (2024).
\newblock Better than classical? the subtle art of benchmarking quantum machine learning models.
\newblock {\em arXiv preprint arXiv:2403.07059}.

\bibitem[Cong et~al., 2019]{cong2019quantum}
Cong, I., Choi, S., and Lukin, M.~D. (2019).
\newblock Quantum convolutional neural networks.
\newblock {\em Nature Physics}, 15(12):1273--1278.

\bibitem[Gacon et~al., 2021]{gacon2021simultaneous}
Gacon, J., Zoufal, C., Carleo, G., and Woerner, S. (2021).
\newblock Simultaneous perturbation stochastic approximation of the quantum fisher information.
\newblock {\em Quantum}, 5:567.

\bibitem[Gandelsman et~al., 2022]{gandelsman2022test}
Gandelsman, Y., Sun, Y., Chen, X., and Efros, A. (2022).
\newblock Test-time training with masked autoencoders.
\newblock {\em Advances in Neural Information Processing Systems}, 35:29374--29385.

\bibitem[Havl{\'\i}{\v{c}}ek et~al., 2019]{havlivcek2019supervised}
Havl{\'\i}{\v{c}}ek, V., C{\'o}rcoles, A.~D., Temme, K., Harrow, A.~W., Kandala, A., Chow, J.~M., and Gambetta, J.~M. (2019).
\newblock Supervised learning with quantum-enhanced feature spaces.
\newblock {\em Nature}, 567(7747):209--212.

\bibitem[Herrmann et~al., 2022]{herrmann2022realizing}
Herrmann, J., Llima, S.~M., Remm, A., Zapletal, P., McMahon, N.~A., Scarato, C., Swiadek, F., Andersen, C.~K., Hellings, C., Krinner, S., et~al. (2022).
\newblock Realizing quantum convolutional neural networks on a superconducting quantum processor to recognize quantum phases.
\newblock {\em Nature communications}, 13(1):4144.

\bibitem[Huang et~al., 2021]{huang2021power}
Huang, H.-Y., Broughton, M., Mohseni, M., Babbush, R., Boixo, S., Neven, H., and McClean, J.~R. (2021).
\newblock Power of data in quantum machine learning.
\newblock {\em Nature communications}, 12(1):2631.

\bibitem[Huang and Goan, 2024]{huang2024l2o}
Huang, Y.-C. and Goan, H.-S. (2024).
\newblock L2{O}-g†: Learning to {O}ptimize {P}arameterized {Q}uantum {C}ircuits with {F}ubini-{S}tudy {M}etric {T}ensor.
\newblock {\em arXiv preprint arXiv:2407.14761}.

\bibitem[Kendall et~al., 2018]{kendall2018multi}
Kendall, A., Gal, Y., and Cipolla, R. (2018).
\newblock Multi-task learning using uncertainty to weigh losses for scene geometry and semantics.
\newblock In {\em Proceedings of the IEEE conference on computer vision and pattern recognition}, pages 7482--7491.

\bibitem[Larocca et~al., 2023]{larocca2023theory}
Larocca, M., Ju, N., Garc{\'\i}a-Mart{\'\i}n, D., Coles, P.~J., and Cerezo, M. (2023).
\newblock Theory of overparametrization in quantum neural networks.
\newblock {\em Nature Computational Science}, 3(6):542--551.

\bibitem[Liu et~al., 2021]{liu2021ttt}
Liu, Y., Kothari, P., Van~Delft, B., Bellot-Gurlet, B., Mordan, T., and Alahi, A. (2021).
\newblock Ttt++: When does self-supervised test-time training fail or thrive?
\newblock {\em Advances in Neural Information Processing Systems}, 34:21808--21820.

\bibitem[Mari et~al., 2020]{mari2020transfer}
Mari, A., Bromley, T.~R., Izaac, J., Schuld, M., and Killoran, N. (2020).
\newblock Transfer learning in hybrid classical-quantum neural networks.
\newblock {\em Quantum}, 4:340.

\bibitem[Mottonen et~al., 2004]{mottonen2004transformation}
Mottonen, M., Vartiainen, J.~J., Bergholm, V., and Salomaa, M.~M. (2004).
\newblock Transformation of quantum states using uniformly controlled rotations.
\newblock {\em arXiv preprint quant-ph/0407010}.

\bibitem[P{\'e}rez-Salinas et~al., 2020]{perez2020data}
P{\'e}rez-Salinas, A., Cervera-Lierta, A., Gil-Fuster, E., and Latorre, J.~I. (2020).
\newblock Data re-uploading for a universal quantum classifier.
\newblock {\em Quantum}, 4:226.

\bibitem[Romero et~al., 2017]{romero2017quantum}
Romero, J., Olson, J.~P., and Aspuru-Guzik, A. (2017).
\newblock Quantum autoencoders for efficient compression of quantum data.
\newblock {\em Quantum Science and Technology}, 2(4):045001.

\bibitem[Shor, 1997]{Shor_1997}
Shor, P.~W. (1997).
\newblock Polynomial-time algorithms for prime factorization and discrete logarithms on a quantum computer.
\newblock {\em SIAM Journal on Computing}, 26(5):1484–1509.

\bibitem[Snell et~al., 2024]{snell2024scaling}
Snell, C., Lee, J., Xu, K., and Kumar, A. (2024).
\newblock Scaling llm test-time compute optimally can be more effective than scaling model parameters.
\newblock {\em arXiv preprint arXiv:2408.03314}.

\bibitem[Stokes et~al., 2020]{stokes2020quantum}
Stokes, J., Izaac, J., Killoran, N., and Carleo, G. (2020).
\newblock Quantum natural gradient.
\newblock {\em Quantum}, 4:269.

\bibitem[Sun et~al., 2024]{sun2024learning}
Sun, Y., Li, X., Dalal, K., Xu, J., Vikram, A., Zhang, G., Dubois, Y., Chen, X., Wang, X., Koyejo, S., et~al. (2024).
\newblock Learning to (learn at test time): Rnns with expressive hidden states.
\newblock {\em arXiv preprint arXiv:2407.04620}.

\bibitem[Sun et~al., 2020]{sun2020test}
Sun, Y., Wang, X., Liu, Z., Miller, J., Efros, A., and Hardt, M. (2020).
\newblock Test-time training with self-supervision for generalization under distribution shifts.
\newblock In {\em International conference on machine learning}, pages 9229--9248. PMLR.

\bibitem[Wang et~al., 2020]{wang2020tent}
Wang, D., Shelhamer, E., Liu, S., Olshausen, B., and Darrell, T. (2020).
\newblock Tent: Fully test-time adaptation by entropy minimization.
\newblock {\em arXiv preprint arXiv:2006.10726}.

\bibitem[Wang et~al., 2023]{wang2023test}
Wang, R., Sun, Y., Gandelsman, Y., Chen, X., Efros, A.~A., and Wang, X. (2023).
\newblock Test-time training on video streams.
\newblock {\em arXiv preprint arXiv:2307.05014}.

\bibitem[Wei et~al., 2022]{wei2022quantum}
Wei, S., Chen, Y., Zhou, Z., and Long, G. (2022).
\newblock A quantum convolutional neural network on nisq devices.
\newblock {\em AAPPS Bulletin}, 32:1--11.

\bibitem[Wen et~al., 2024]{wen2024enhancing}
Wen, J., Huang, Z., Cai, D., and Qian, L. (2024).
\newblock Enhancing the expressivity of quantum neural networks with residual connections.
\newblock {\em Communications Physics}, 7(1):220.

\end{thebibliography}
\bibliographystyle{apalike}

\newpage  %
\normalsize
\titlespacing*{\section}{0pt}{*1}{*1}
\titlespacing*{\subsection}{0pt}{*1.25}{*1.25}
\titlespacing*{\subsubsection}{0pt}{*1.5}{*1.5}

\setlength{\abovedisplayskip}{10pt}
\setlength{\abovedisplayshortskip}{10pt}
\setlength{\belowdisplayskip}{10pt}
\setlength{\belowdisplayshortskip}{10pt}

\setlist[itemize]{leftmargin=1em, before=\vspace{-0.2em}, after=\vspace{-0.2em}, itemsep=0.1em}
\setlist[enumerate]{leftmargin=1.4em, 
before=\vspace{-0.2em}, after=\vspace{-0.2em}, 
itemsep=0.1em}

\end{document}